\documentclass[times,review,preprint,authoryear]{elsarticle}

\usepackage{medima}
\usepackage{framed,multirow}
\usepackage{makecell}

\usepackage{amssymb}
\usepackage{latexsym}
\usepackage[labelfont=bf]{caption}
\usepackage{subcaption}
\usepackage{url}
\usepackage{xcolor}
\usepackage{svg}
\usepackage{amsmath}
\usepackage[hidelinks]{hyperref}
\usepackage{graphicx}
\usepackage{tabularx,booktabs}
\definecolor{newcolor}{rgb}{.8,.349,.1}

\journal{Medical Image Analysis}

\begin{document}

\verso{Given-name Surname \textit{et~al.}}

\begin{frontmatter}

\title{TotalRegistrator: Towards a Lightweight Foundation Model for CT Image Registration}
%\tnoteref{tnote1}}%
%\tnotetext[tnote1]{This is an example for title footnote coding.}

% \author[1,2]{Xuan Loc \snm{Pham}\corref{cor1}}
\author[1]{Xuan Loc \snm{Pham}}

\author[1]{Gwendolyn \snm{Vuurberg}}
\author[1]{Marjan \snm{Doppen}}
\author[1]{Joey \snm{Roosen}}
\author[1]{Tip \snm{Stille}}

\author[5]{Thi Quynh \snm{Ha}}
\author[6]{Thuy Duong \snm{Quach}}
\author[5]{Quoc Vu \snm{Dang}}

\author[7]{Manh Ha \snm{Luu}}

\author[1]{Ewoud J. \snm{Smit}}
\author[4]{Hong Son \snm{Mai}}
\author[3]{Mattias \snm{Heinrich}}
\author[1,2]{Bram \snm{van Ginneken}}
\author[1]{Mathias \snm{Prokop}}

\author[1]{Alessa \snm{Hering}\corref{cor1}}
\cortext[cor1]{Corresponding author at: Department of Imaging, Radboudumc, Nijmegen, the Netherlands;}
\ead{alessa.hering@radboudumc.nl}

\address[1]{Department of Imaging, Radboudumc, Nijmegen, the Netherlands}
\address[2]{Fraunhofer MEVIS, Bremen, Germany}
\address[3]{Institute for Medical Informatics, University of Lübeck, Lübeck, Germany}
\address[4]{Department of Nuclear Medicine, Hospital 108, Hanoi, Vietnam}
\address[5]{Department of Diagnostic Imaging and Interventional Radiology, Thai Nguyen National Hospital, Thai Nguyen, Vietnam}
\address[6]{Diagnostic Imaging and Interventional Radiology Center, Tam Anh Hospital, Hanoi, Vietnam}
\address[7]{FET, Vietnam National University, University of Engineering and Technology, Hanoi, Vietnam}

% \received{1 May 2022}
% \finalform{10 May 2022}
% \accepted{13 May 2022}
% \availableonline{15 May 2022}
% \communicated{Xuan Loc Pham}

\begin{abstract}
%%%
Image registration is a fundamental technique in the analysis of longitudinal and multi-phase CT images within clinical practice. However, most existing methods are tailored for single-organ applications, limiting their generalizability to other anatomical regions. A general-purpose registration model would streamline clinical pipelines and reduce the need for task-specific models, which are often resource-intensive to train and maintain. While a few models have addressed registration across the thorax and abdomen, broadly applicable approaches remain limited. 

This work presents \textit{TotalRegistrator}, an image registration framework capable of aligning multiple anatomical regions simultaneously using a standard UNet architecture and a novel field decomposition strategy. The model is lightweight, requiring only 11GB of GPU memory for training. To train and evaluate our method, we constructed a large-scale longitudinal dataset comprising 695 whole-body (thorax-abdomen-pelvic) paired CT scans from individual patients acquired at different time points. 
We benchmarked TotalRegistrator against a generic classical iterative algorithm and a recent foundation model for image registration. To further assess robustness and generalizability, we evaluated our model on three external datasets: the public thoracic and abdominal datasets from the Learn2Reg challenge, and a private multiphase abdominal dataset from a collaborating hospital.
Experimental results on the in-house dataset show that the proposed approach generally surpasses baseline methods in multi-organ abdominal registration, with a slight drop in lung alignment performance. On out-of-distribution datasets, it achieved competitive results compared to leading single-organ models, despite not being fine-tuned for those tasks, demonstrating strong generalizability. The source code will be publicly available prior to publication at: \url{https://github.com/DIAGNijmegen/oncology_image_registration.git}.

\end{abstract}


\end{frontmatter}

% \linenumbers

\section{Introduction} \label{Intro}
\begin{figure} [ht]
\begin{center}
\includegraphics[width=0.95\textwidth]{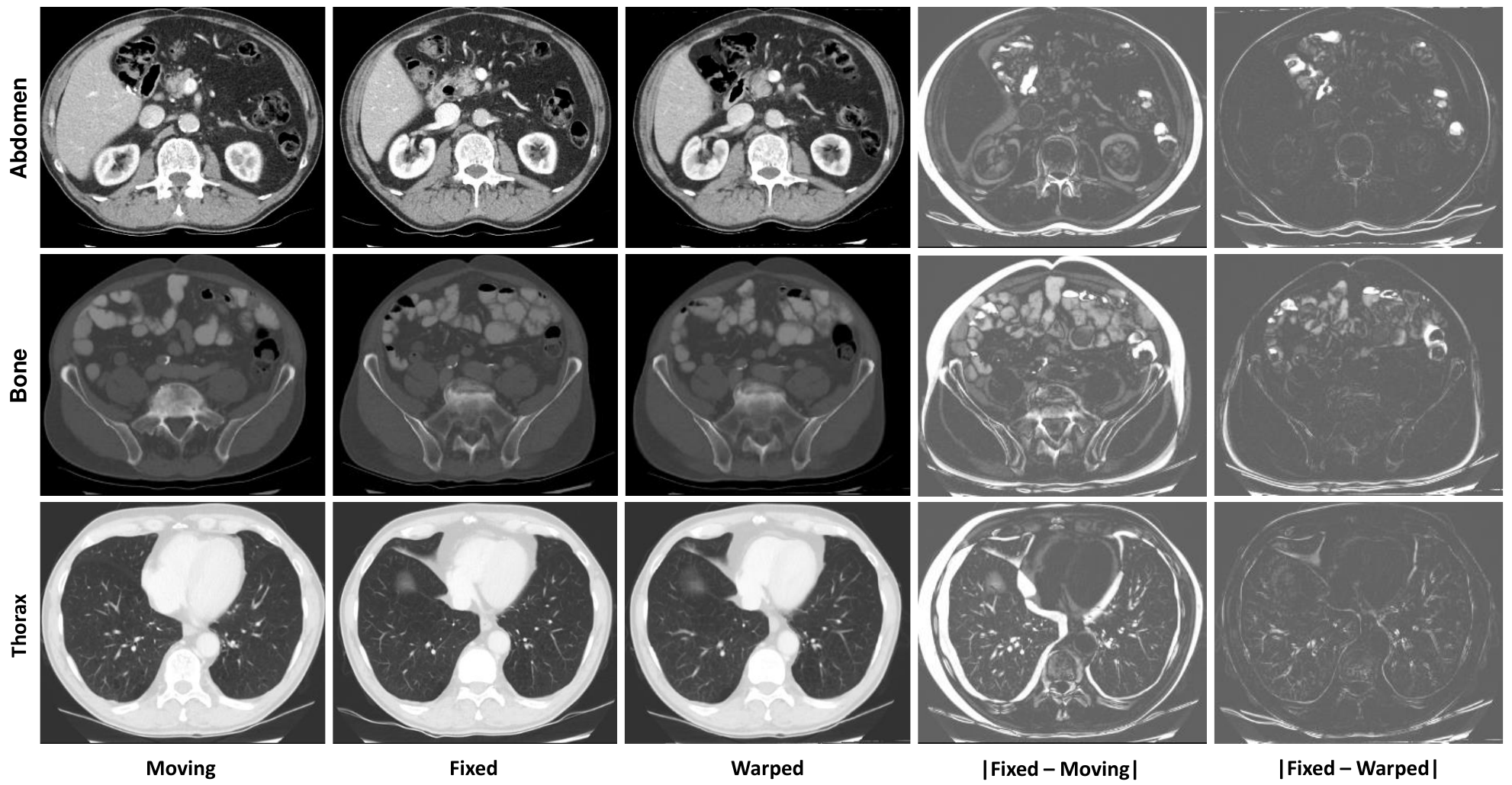}
\end{center}
\caption[example] 
{ \label{fig:intro_totalreg} 
Illustration of the multi-organ registration by TotalRegistrator in different regions. The first row shows an example case of the abdomen region, while the next two rows illustrate cases of the bone and thorax regions, respectively. The last two columns indicate the absolute intensity subtraction before and after alignment to verify the registration results.   
}
\end{figure} 

Image registration plays a crucial role in the cancer treatment pipeline, supporting tasks such as treatment planning, image-guided intervention, and treatment follow-up \citep{Hering24, Luu23, Gunay17}. For example, aligning diagnostic and intra-operative CT scans can improve navigation during surgery~\citep{Gunay17}, while longitudinal registration enables accurate lesion assessment over time~\citep{Hering24}. Despite its wide applicability, current registration methods are typically designed for specific organs or tasks, limiting their use in workflows that span multiple anatomical regions. This fragmentation is particularly problematic in cancer care, where longitudinal and whole-body imaging is common. A general-purpose registration model capable of aligning diverse anatomical structures with low latency could simplify clinical integration and reduce the need for task-specific models.

Traditional image registration methods are based on iterative optimization, which progressively refines the alignment between two images by minimizing a similarity metric. These approaches, such as SyN~\citep{Avants08}, Elastix~\citep{Klein10}, and ANTs~\citep{Avants14}, are valued for their accuracy and flexibility across different organs and imaging conditions. However, they are computationally intensive, often requiring several minutes per image pair, which limits their applicability in time-sensitive clinical settings. Moreover, performance depends heavily on parameter tuning, such as pyramid resolution, grid spacing, or regularization, which must often be adapted to each new dataset or anatomical region. While recent efforts have accelerated optimization pipelines, e.g., through GPU parallelization~\citep{Budelmann19} or hybrid feature-driven methods~\citep{Siebert25}, these solutions still face challenges in scaling to diverse anatomical structures or rare imaging scenarios. This has driven growing interest in learning-based models, which aim to eliminate the need for per-pair optimization entirely.

Deep learning has enabled a paradigm shift in image registration by replacing iterative optimization with direct prediction of deformation fields. Early approaches such as RegNet~\citep{Sokooti17, Sokooti19} relied on supervised learning, using synthetic or algorithm-generated deformation fields as labels. However, generating sufficient high-quality ground truth deformations proved challenging. This limitation was addressed by VoxelMorph~\citep{Balakrishnan19}, which introduced an unsupervised framework that learns to align image pairs without requiring annotated deformations. Building on this idea, many follow-up methods have adopted pyramid-based strategies, either through cascaded registration blocks~\citep{Shengyu19, Zou21, Pham24} or through single-block architectures capable of multi-scale processing~\citep{Mok20, Hering21, Kang22, Chen22, Haiqiao24}, to mimic the robustness of classical algorithms. Despite these advances, most learning-based methods remain highly task-specific. They are typically trained and optimized for a single anatomical region, and cannot generalize across organs without retraining or architectural modifications. This limits their clinical utility in settings where whole-body or multi-organ registration is required.

A recent step towards general-purpose image registration was introduced by~\cite{Lin24}, who proposed uniGradICON, a preliminary foundation model trained across diverse anatomical regions and imaging modalities. Built upon the gradient inverse consistency regularizer (GradICON)~\citep{Lin23}, the model was trained using a mixture of public datasets, including brain, lung, liver, and knee scans from CT and MRI modalities. Remarkably, the same architecture and hyperparameter set were used across all training tasks, yielding competitive performance compared to top-performing single-organ models. However, uniGradICON’s reliance on heterogeneous datasets with imbalanced anatomical coverage may limit its robustness, particularly in clinically complex regions like the abdomen. Moreover, the method requires substantial computational resources due to the cost of the GradICON regularization and the scale of the training data. As such, the model’s design aligns with the traditional paradigm of foundation models — highly performant but difficult to train and deploy in settings with constrained computing resources, such as many hospitals or smaller research institutions.

In this paper, we propose \textit{TotalRegistrator}, a general-purpose unsupervised learning-based model for whole-body CT image registration (Fig. \ref{fig:intro_totalreg}). Unlike prior approaches that rely on diverse public datasets or extensive computing resources, the proposed method is trained from scratch using a single, carefully curated longitudinal dataset and is fully trainable on standard, low-end GPUs. We demonstrate that sufficient anatomical diversity for training a broadly applicable model can be captured from intra-patient scans acquired across time and anatomical regions. This offers a practical alternative to large-scale dataset aggregation while significantly lowering hardware requirements. A key challenge in multi-organ registration lies in simultaneously modeling deformations across organs with widely varying size, shape, and location (e.g., pancreas vs. liver, prostate vs. lung). To address this, we introduce a field decomposition strategy inspired by multi-cascade architectures~\citep{Shengyu19, Zou21, Pham24}, in which region-specific registration blocks independently estimate local deformations for specific groups of adjacent anatomical structures. This field decomposition approach helps concentrate spatially related organs from a specific region, thereby significantly reducing the complexity of the collective deformation fields. These local deformations are then fused by a final integration block to produce a coherent whole-body deformation field. To ensure accessibility and reproducibility, we adopt a minimal design based on the standard UNet architecture~\citep{Ronneberger15} for registration blocks, making the model easy to train and deploy using conventional hardware.

The preliminary version of TotalRegistrator was introduced at the SPIE Medical Imaging 2025 conference~\citep{Pham25}. In this work, we present a substantially extended version, with the following contributions:        
\begin{itemize}
    \item \textbf{Field decomposition for multi-region registration:} We introduce a novel field decomposition strategy to support efficient and accurate registration across anatomically diverse regions. Compared to the preliminary version, we expand the model to include a dedicated bone registration block alongside the thoracic and abdominal components. This addition increases anatomical coverage and improves the model’s robustness to complex deformation patterns.
    
    \item \textbf{A lightweight, general-purpose CT registration model:} TotalRegistrator is one of the first general models for CT image registration that is lightweight, easily implementable, and broadly applicable. The multi-region training strategy is designed to run on standard 11GB GPUs. We further refine the final whole-body block to handle potential folding artifacts arising from the sequential integration of multiple regional deformation fields.

    \item \textbf{A curated, large-scale longitudinal CT dataset:} We construct a dataset of 695 longitudinal intra-patient thorax-abdomen-pelvis CT scan pairs. Among these, 104 cases are manually annotated by multiple clinical experts covering 12 anatomical structures, representing a significant improvement over the preliminary version, which included only four organs with automatically generated labels.
\end{itemize}

% They focus on the traditional way to create a foundation model: big datasets from around the world + high-end HPC. 

\section{Methods} \label{Methods}
\subsection{Problem Formulation}
A volumetric image registration task aims to determine the optimal spatial alignment between two input images, as formulated in equations \eqref{IW}, \eqref{Q}. In this context, \( I_F \), \( I_M \), and \( I_W \) denote the fixed, moving, and warped images, respectively, within the spatial domain \(\Omega \subset \mathbb{R}^3\). The transformation field \(\Phi\) defines the mapping that aligns \( I_M \) and \( I_F \).
\begin{equation}
    I_W = I_M \circ \Phi \quad,
\label{IW}
\end{equation}
\begin{equation}
    \hat{\Phi}=\arg{\min_\Phi{\mathcal{L}\left(I_F,I_M,\Phi\right)}} \quad,
\label{Q}
\end{equation}
In single-organ registration scenarios, existing deep learning-based methods primarily focus on optimizing a single deformation field $\Phi$ such that the organ shape as well as its inner structures in \( I_W \) are brought closest to \( I_F \). In multi-organ registration, although the model can still predict a single deformation field from a larger field of view (e.g., thorax-abdomen), the presence of heterogeneous anatomical structures introduces greater modeling complexity. This arises from the need to capture and balance different types of tissue motion and anatomical relationships within a single deformation field. While this does not inherently increase the computational cost beyond that imposed by the input image size, it does make the learning task more challenging. In this study, this extension of \(\Phi\) is defined in equation \eqref{O_compose}.
\begin{equation}
    \Phi = \sum_{structure=1}^{total} \Phi_{\text{structure}}
\label{O_compose}
\end{equation}

\subsection{TotalRegistrator}
\begin{figure} [ht]
\begin{center}
\begin{tabular}{c} 
\includegraphics[width=0.8\textwidth]{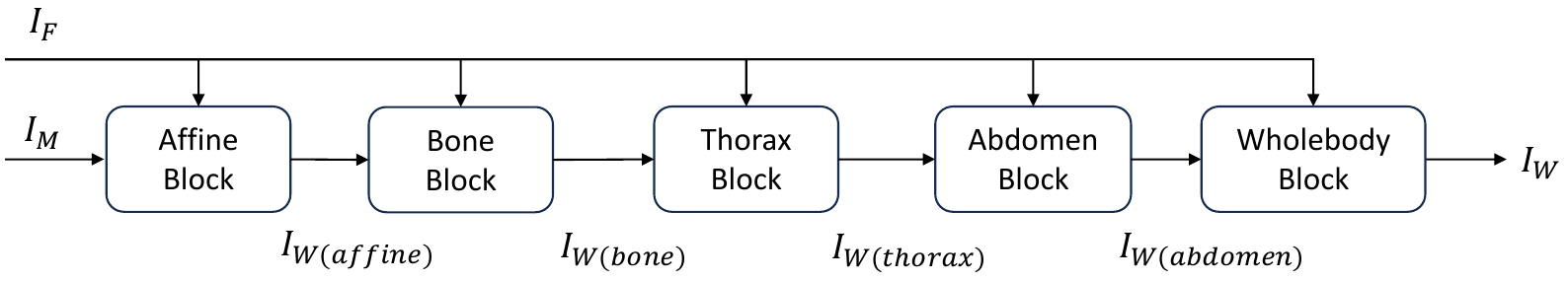}
\end{tabular}
\end{center}
\caption[architecture] 
{ \label{fig:pipeline_diagram} 
Illustration of the proposed field decomposition technique.
}
\end{figure} 

In this study, we propose a reinterpretation of equation \eqref{O_compose} through a task-based and region-specific field decomposition approach (Fig. \ref{fig:pipeline_diagram}) to effectively manage the complexity inherent in multi-organ registration. As shown in equation \eqref{O_propose}, the compositional deformation field \(\Phi\) is explicitly partitioned into \(\Phi_{affine}\) and \(\Phi_{deformable}\) components. The deformable registration component is further subdivided to account for region-specific deformations within the thoracic, abdominal, and skeletal structures. Finally, a global deformation field \(\Phi_{wholebody}\) is introduced to refine and harmonize the contributions of these region-specific components, thereby ensuring a cohesive and anatomically consistent alignment across the entire body.
\begin{align}
    \Phi &= \Phi_{affine} + \Phi_{deformable} \notag \\
         &= \Phi_{affine} + \Phi_{bone} + \Phi_{thorax} + \Phi_{abdomen} + \Phi_{wholebody}.
\label{O_propose}
\end{align}
Equation \eqref{L} defines the loss function employed to optimize the field $\Phi_{deformable}$. This function comprises three main components: the Mutual Information (MI) loss $\mathcal{L}_{MI}$ \citep{Thevenaz00} for quantifying image similarity; the Dice Similarity Coefficient (DSC) loss $\mathcal{L}_{DSC}$ for assessing segmentation overlap; and the Bending Energy (BE) loss $\mathcal{L}_{BE}$ serving as a regularization term to enforce smoothness in the deformation field. $S_F$ and $S_W$ denote the segmentation masks for the fixed and warped images, $L_F$ and $L_M$ represent sets of discrete intensity levels in the fixed and moving images, while $p$ is the discrete joint probability. The weighting factors $\alpha$, $\lambda$, and $\beta$ are introduced to balance the contributions of these loss terms to the overall optimization process.
\begin{align}
\mathcal{L}\left(I_F,I_M,\Phi_{deformable}\right) &= \alpha\mathcal{L}_{MI}\left(I_F,I_W\right)+\lambda\mathcal{L}_{DSC}\left(S_F,S_W\right)+\beta\mathcal{L}_{BE}\left(\Phi\right)  \notag \\
&=
    -\alpha \sum_{m \in L_M} \sum_{f \in L_F}
p(f, m; \Phi) \log_2 \frac{p(f, m; \Phi)}{p_F(f) \cdot p_M(m; \Phi)}
    + \lambda \left( 1 - \frac{2 \sum_{\mathbf{x} \in \Omega} S_F(\mathbf{x}) S_W(\mathbf{x})}{\sum_{\mathbf{x} \in \Omega} S_F(\mathbf{x}) + \sum_{\mathbf{x} \in \Omega} S_W(\mathbf{x})} \right) \notag \\
&\quad + \beta \frac{1}{|\Omega|} \sum_{\mathbf{x} \in \Omega} \sum_{k=1}^{3} \left\| \nabla^2 \Phi_k(\mathbf{x}) \right\|_F^2.
\label{L}
\end{align}

% \begin{equation}
% \mathcal{L}\left(I_F,I_M,\Phi\right)=\alpha\mathcal{L}_{MI}\left(I_F,I_W\right)+\lambda\mathcal{L}_{DSC}\left(S_F,S_W\right)+\beta\mathcal{L}_{BE}\left(\Phi\right),
% \end{equation}
% The MI loss measures the statistical dependency between the fixed image \(I_F\) and the warped image \(I_W\):
% \begin{equation}
%     \mathcal{L}_{MI}\left(I_F, I_W\right) = -\sum_{i,j} p_{I_F, I_W}(i,j) \log \frac{p_{I_F, I_W}(i,j)}{p_{I_F}(i) p_{I_W}(j)},
% \end{equation}
% where \( p_{I_F, I_W}(i,j) \) is the joint probability distribution of \( I_F \) and \( I_W \), while \( p_{I_F}(i) \) and \( p_{I_W}(j) \) are their marginal probability distributions.
% The DSC loss evaluates the overlap between the segmentation masks \( S_F \) and \( S_W \):
% \begin{equation}
%     \mathcal{L}_{DSC}\left(S_F, S_W\right) = 1 - \frac{2 \sum_{\mathbf{x} \in \Omega} S_F(\mathbf{x}) S_W(\mathbf{x})}{\sum_{\mathbf{x} \in \Omega} S_F(\mathbf{x}) + \sum_{\mathbf{x} \in \Omega} S_W(\mathbf{x})}.
% \end{equation}
% The BE loss penalizes excessive curvature in the transformation field \( \Phi \), enforcing smooth deformations:
% \begin{equation}
%     \mathcal{L}_{BE}(\Phi) = \sum_{\mathbf{x} \in \Omega} \sum_{i,j} \left( \frac{\partial^2 \Phi}{\partial x_i \partial x_j} \right)^2.
% \end{equation}
% Here, \( \Phi \) denotes the deformation field, and the second-order derivatives ensure smoothness by penalizing high bending energy.

In equation \eqref{L_affine}, we use the same loss function as in Equation \eqref{L}. However, the penalty term is omitted during the optimization of $\Phi_{\text{affine}}$, since the registration task lacks abrupt intensity variations that might induce unrealistic deformations.
\begin{equation}
\mathcal{L}\left(I_F,I_M,\Phi_{affine}\right)=\alpha\mathcal{L}_{MI}\left(I_F,I_W\right)+\lambda\mathcal{L}_{DSC}\left(S_F,S_W\right)
\label{L_affine}
\end{equation}

\subsection{Model Architecture}
\subsubsection{Base Registration Block}

\begin{figure} [ht]
\begin{center}
\begin{tabular}{c} 
\includegraphics[width=0.9\textwidth]{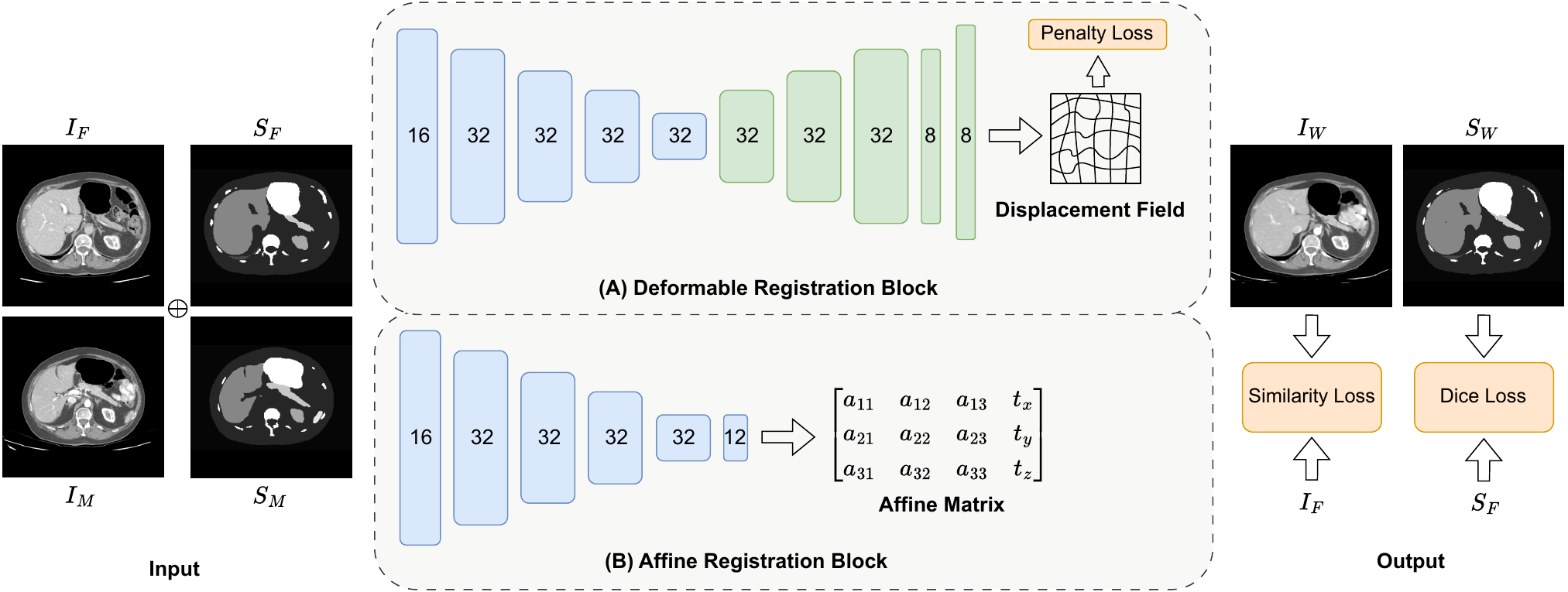}
\end{tabular}
\end{center}
\caption[architecture1] 
{ \label{fig:baseblock} 

Illustration of the base architectures for the affine and deformable registration blocks. The deformable registration blocks use the standard encoder-decoder architecture for feature learning and generating the dense displacement field. Meanwhile, the affine block only relies on the encoder branch to output 12 parameters for the affine transform matrix. Pairs of input images and multi-organ segmentation masks are provided during the training phase. The unsupervised learning process is managed by a combination of similarity loss, segmentation overlay loss, and deformation field regularization loss (not used in training affine registration). During inference, only pairs of input images are needed without the multi-organ segmentation masks.   
}
\end{figure} 

TotalRegistrator is constructed from multiple registration blocks with the same UNet-like architecture. The affine block is composed exclusively of the encoder part, with its final layer generating a tensor representing the 12 degrees of freedom for the affine transformation. Meanwhile, the deformable registration blocks adopt a comprehensive encoder-decoder architecture. Details of the model architecture are illustrated as in Figure \ref{fig:baseblock}.

The encoder side consists of five stages with the number of channels of 16, 32, 32, 32, and 32, respectively. Through each stage, image features are extracted and encoded gradually until the coarsest final layer. In the first stage, we utilize a 3D convolutional layer with a minimum stride of 1, kernel size 3 and padding 1 to convert the number of input channels to 16 while preserving the image spatial dimension. In the next stages, we increase the stride to 2 to downsample the image size by half after each stage. The 3D convolution is then followed by a Leaky ReLU function with a threshold value of 0.2 for better gradient control during training. 

The decoder path includes six stages with the number of channels of 32, 32, 32, 8, 8, and 3, respectively. In each decoding stage, we first upsample the output from the previous stage using a 3D transposed convolutional layer with stride 2, padding 1, and kernel size 3. To preserve the spatial information of image features in the decoding phase, we also pass the encoding information from the corresponding resolution of the encoding side to the decoding side by a skip connection path. This skip-connected encoding information is subsequently concatenated with the upsampled decoding information to become the input of the current decoding stage. The concatenated feature maps are then passed through the same series of 3D convolution and Leaky ReLU layers as the encoder side. The final decoding layer is responsible for generating feature maps for the output deformation field with three channels. Ultimately, the generated deformation field is warped with the input moving image by a Spatial Transformation Network (STN)~\citep{Jaderberg15} to form the warped image.

\subsubsection{Field Decomposition Apporach}

\begin{figure} [ht]
\begin{center}
\begin{tabular}{c} 
\includegraphics[width=0.95\textwidth]{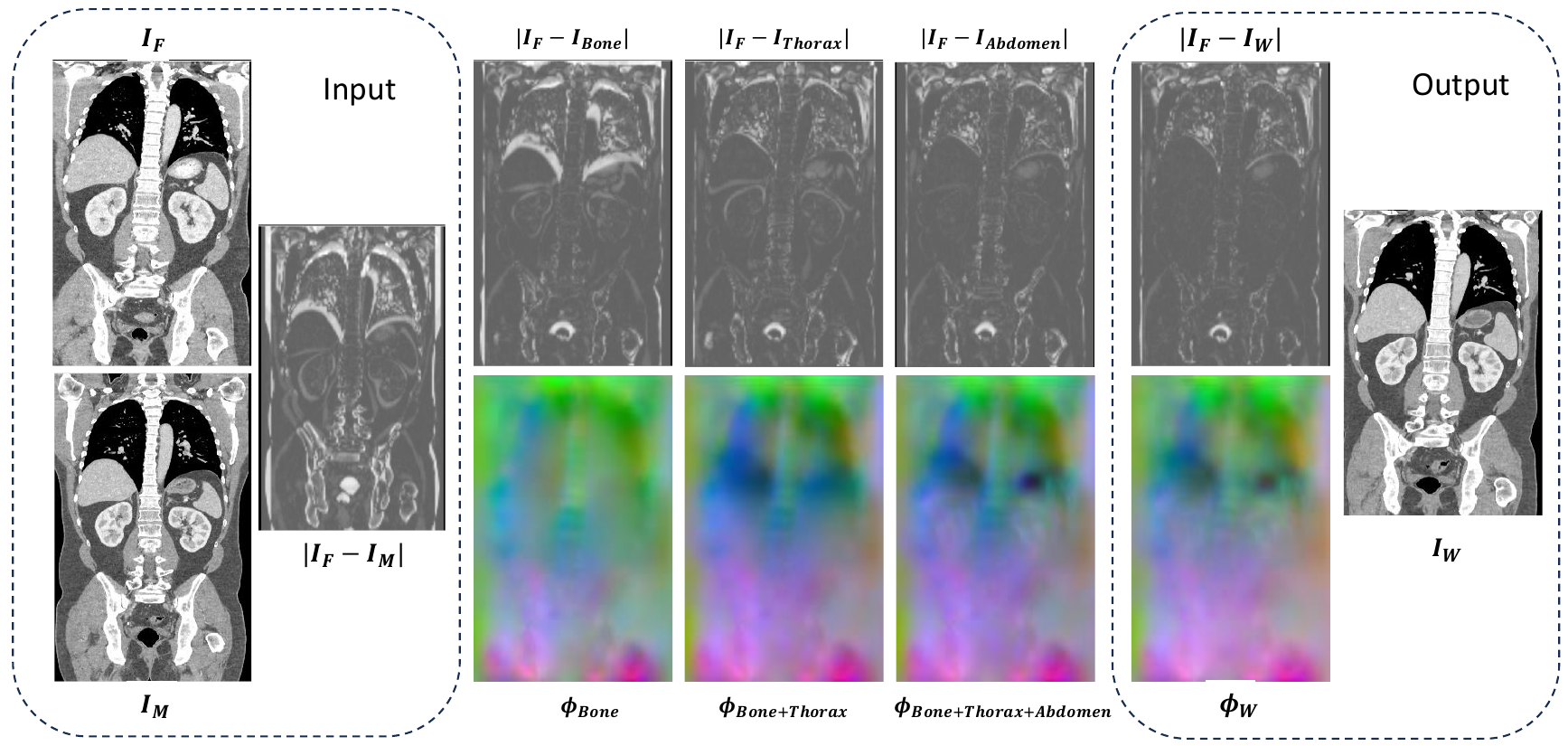}
\end{tabular}
\end{center}
\caption[architecture2] 
{ \label{fig:pipeline} 
Illustration of the task-based region-specific field decomposition approach. The visualization of each displacement field indicates the region-of-focus of each registration block. The absolute difference row further demonstrates the deformations after each block. The final displacement field is formed by the accumulation of all preceding displacements. 
}
\end{figure} 

Figure \ref{fig:pipeline} illustrates the field decomposition approach proposed in equation \eqref{O_propose}. To mitigate the complexity of the registration task, we employ distinct registration blocks for affine and deformable registration. Given the straightforward nature of affine registration, a single block is sufficient for this task. In contrast, multi-organ deformable registration necessitates the simultaneous optimization of multiple deformation fields corresponding to anatomical structures with varying shapes, sizes, and spatial locations. To address this challenge, we partition the whole-body view into three specific regions: the thoracic, abdominal, and bone regions. Each region is assigned an independent deformable registration block trained exclusively on the corresponding anatomical structures, guided by segmentation masks. Specifically, the thoracic block registers organs such as the heart and lungs, the abdominal block handles organs including the liver, kidneys, pancreas, stomach, spleen and gallbladder, while the bone block processes skeletal structures such as the pelvis, vertebrae, and ribs. This clusterization technique reduces complexity by allowing each registration block to focus on localized image features. Moreover, adjacent organs within the same region often exhibit structural similarities, facilitating a more efficient and effective learning process.        

Rather than adopting the widely used recursive registration training strategy~\citep{Shengyu19, Pham24}, which integrates the entire pipeline into a single training framework, we train each registration block separately and independently. This approach ensures the model remains feasible for training on standard machines with limited computational resources. Furthermore, rather than maintaining a single large cumulative gradient graph shared across all anatomical structures, we propose performing the forward and backward propagation cycle separately for each organ. For instance, in each training epoch of the abdominal block, instead of executing one backward pass that stores six gradient graphs for six organs, we perform six individual backward passes, each corresponding to a single organ. By allocating memory only for the computational graph of a single organ at a time, this strategy significantly reduces the overall memory requirements during training. 

Depending on the deformations' complexity, the field decomposition and composition pipeline follows a predefined order: the affine block, the bone block, the thorax block, the abdomen block and finally the whole-body block to smooth out and balance the impact of other blocks on the overall performance. We begin by training the affine registration block, whose output serves as the foundation for training the subsequent deformable registration blocks. Within the overall pipeline, the deformation fields generated by earlier registration blocks are sequentially combined and applied to the original moving image, producing an updated moving image for the current registration block. For instance, as illustrated in Figure \ref{fig:pipeline}, the accumulated deformations up to the abdomen registration block integrate transformations from preceding stages, such as thoracic and bone deformations. From the preliminary work, we observed that the independently trained whole-body block struggles to manage the incremental foldings caused by the successive addition of multiple deformation fields from previous blocks. The folding-explosion phenomenon is even more pronounced with the introduction of an additional registration block in this study. Therefore, unlike other registration blocks in the pipeline, the whole-body block is trained based on the outputs of the preceding blocks, which are kept frozen. This strategy enables the whole-body block to more effectively anticipate and regularize the levels of folding encountered during inference. 

\section{Experiments and Materials} \label{Experiments}
Experiments in this study involve one in-house dataset and three external datasets as summarized in Table \ref{tab:dataset_summary}. Details regarding those datasets will be discussed in this section.

\begin{table}[ht]
\caption{Summary of datasets used in this study. RUMC is the in-house dataset used for both the training phase and the evaluation of TotalRegistrator. The other three external datasets are used for ablation studies, where AbdomenCTCT and NLST are downloadable from the Learn2Reg challenge (\url{https://learn2reg.grand-challenge.org/Datasets/}), while H108M was obtained from our partner hospital in Vietnam.}

\vspace{10pt}  % Adds additional vertical space between caption and table
\label{tab:dataset_summary}
\centering
\small
\renewcommand{\arraystretch}{2.5} % Increase cell height
\begin{tabular}{|>{\centering\arraybackslash}p{2.8cm}|>{\centering\arraybackslash}p{1.4cm}|>{\centering\arraybackslash}p{1.4cm}|>{\centering\arraybackslash}p{1.8cm}|>{\centering\arraybackslash}p{3.8cm}|}
\hline
\textbf{Dataset} & \textbf{Training (pairs)} & \textbf{Evaluation (pairs)} & \textbf{Region} & \textbf{Type} \\
\hline
\makecell{RUMC \\ (in-house dataset)} & 591 & 104 & Whole body & Intra-patient longitudinal \\
\hline
\makecell{AbdomenCTCT \\ \citep{Xu16}} & N/A & 45 & Abdomen & Inter-patient \\
\hline
\makecell{NLST \\ \citep{Aberle11}} & N/A & 10 & Lung & Intra-patient longitudinal \\
\hline
\makecell{H108M \\ \citep{Pham24}} & N/A & 136 & Liver & Intra-patient multiphase \\
\hline
\end{tabular}
\end{table}

\subsection{Data Acquisition}
\subsubsection{In-house dataset}
In this study, we retrospectively collected a large-scale in-house dataset comprising 695 pairs of whole-body CT scans from patients treated at Radboud University Medical Center Nijmegen (RUMC) between 2011 and 2021. Each pair consists of two scans of the same patient acquired at different time points. All images provide a comprehensive view of the body, covering the thoracic, abdominal, and pelvic regions, with consistent slice thickness within each pair. To minimize potential biases related to demographic and technical factors such as age, gender, and scanner variations, we carefully curated a diverse patient cohort, ranging in age from 17 to 87 years, with a mean age of 58.8 years. The dataset also maintains a relatively balanced gender distribution, with 57$\%$ male and 43$\%$ female patients. As we are a referral hospital, images in our database were acquired with various CT scanners from different manufacturers (including Canon, GE, Philips and Siemens) leading to the wide availability in slice thickness, ranging from 0.8 mm to 3 mm and 4 mm. The in-plane image size is either $512\times512$ or $1024\times1024$, ensuring a comprehensive representation of real-world clinical imaging conditions.

\begin{figure} [!t]
\begin{center}
\includegraphics[width=0.7\textwidth]{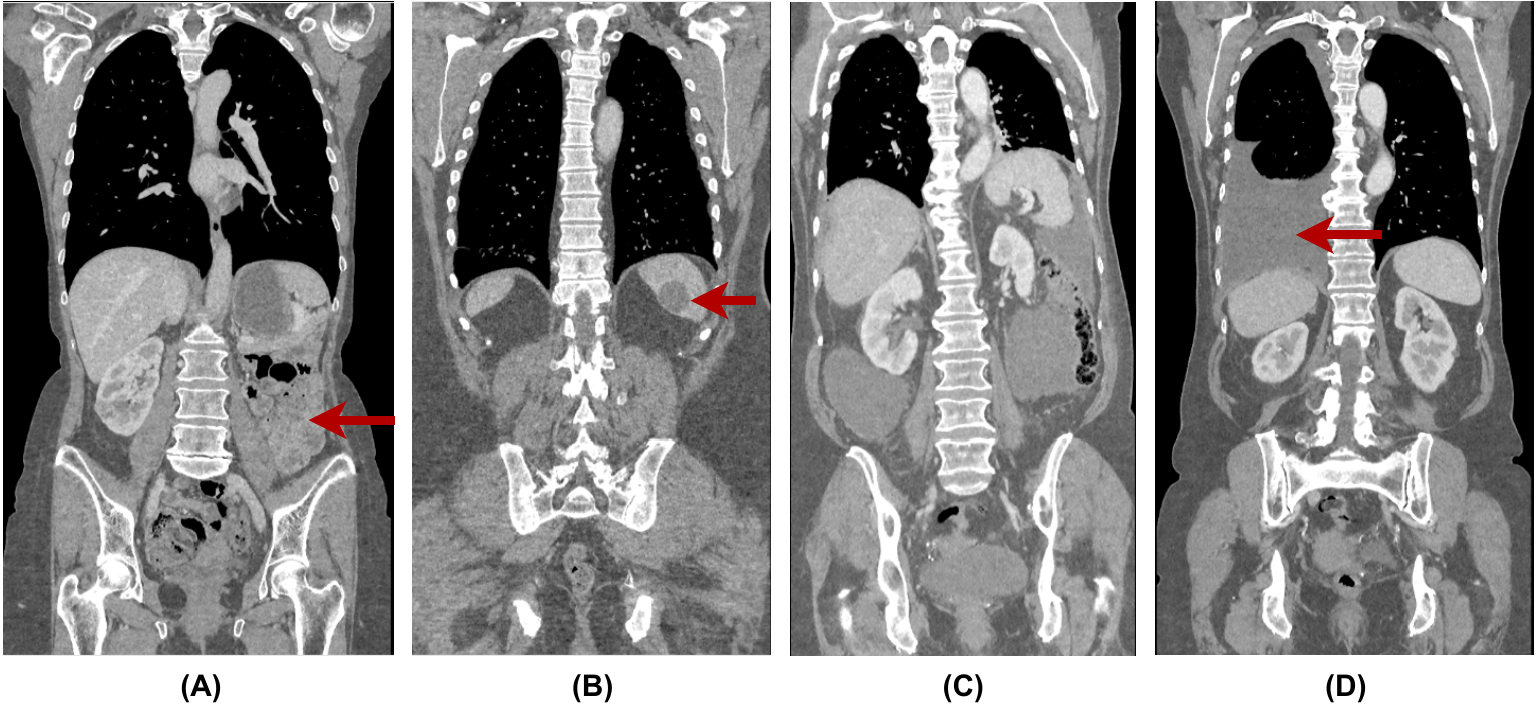}
\end{center}
\caption[dataset] 
{ \label{fig:dataset} Illustration of a case in the in-house Radboudumc dataset with 
(A) post-surgery effects (left kidney removal).
(B) pathological signs (splenic cyst).
(C-D) disease progression between two timepoints (pleural fluid). 
The red arrows point to the region of abnormalities in each image.
}
\end{figure}

The dataset also demonstrates a good range of diversity and variability as most patients underwent some kind of cancer treatment and surgery, making multiple pathological signs present in the dataset. Those with abnormalities can be classified into three types: cases with post-surgery effects, cases with anatomical abnormalities, and cases with disease progression. Representative examples of each category are illustrated in Figure \ref{fig:dataset}. Specifically, Figure \ref{fig:dataset} (A) illustrates a case with common post-surgical effects, including organ nephrectomy and spinal fusion surgery. Figure \ref{fig:dataset} (B) highlights a case with distinct pathological signs, including gastroenterostomy with gastric tube reconstruction, lung cancer and kidney cysts, while Figure~\ref{fig:dataset} (C-D) presents a case of disease progression, such as liver and lung cancer metastasis.

% \begin{figure}[htbp]
%     \centering
%     \begin{subfigure}[b]{0.8\textwidth}
%         \centering
%         \includegraphics[width=\textwidth]{./img/chap4/4_1_a.PNG}
%         \caption{Illustration of cases with post-surgery effects. (A) Lung lobes removal (B) Left kidney removal (C) Spleen removal (D) Spinal fusion.}
%         \label{fig:dataset_a}
%     \end{subfigure}
%     \vspace{5pt} % Adjust vertical space between images
%     \begin{subfigure}[b]{0.8\textwidth}
%         \centering
%         \includegraphics[width=\textwidth]{./img/chap4/4_1_b.PNG}
%         \caption{Illustration of cases with pathological signs. (A) Lung COPD (B) Lung cancer metastasis (C) Kidney cysts (D) Whole body cancer metastasis.}
%         \label{fig:dataset_b}
%     \end{subfigure}
%     \vspace{5pt} % Adjust vertical space between images
%     \begin{subfigure}[b]{0.8\textwidth}
%         \centering
%         \includegraphics[width=\textwidth]{./img/chap4/4_1_c.PNG}
%         \caption{Illustration of cases with disease progression. (A1-A2) Liver metastasis progression (B1-B2) Lung lobe removal due to cancer progression.}
%         \label{fig:dataset_c}
%     \end{subfigure}
%     \caption{Illustration of different types of abnormalities in the dataset. Red arrows point to the noticed abnormalities in each scan.}
%     \label{fig:dataset}
% \end{figure}

\subsubsection{External datasets}
The two public external datasets, AbdomenCTCT~\citep{Xu16} and NLST~\citep{Aberle11}, can be downloaded from the Learn2Reg challenge\footnote{https://learn2reg.grand-challenge.org/Datasets/}~\citep{Hering23}. As the official test data is not made publicly available by challenge organizers, we followed the uniGradICON paper~\citep{Lin24} to utilize the validation subsets for experiments in this study. The AbdomenCTCT dataset contains abdominal CT images from different patients with annotations for 13 organs. The test set was created by following the provided JSON file to form 45 inter-patient pairs. Meanwhile, the NLST test set includes 10 pairs of longitudinal intra-patient lung CT scans, each with lung masks and keypoints.  

The other dataset, H108M, was obtained in our partner hospital in Vietnam~\citep{Pham24}. The dataset includes 136 pairs of multi-phase liver CT images. Each pair contains scans from the same patient, taken at the same time point but in different phases such as the Arterial phase, Portal Venous phase and Delayed phase.  

\subsection{Data Preprocessing}  
\subsubsection{In-house dataset}

\begin{figure} [!t]
\begin{center}
\includegraphics[width=0.95\textwidth]{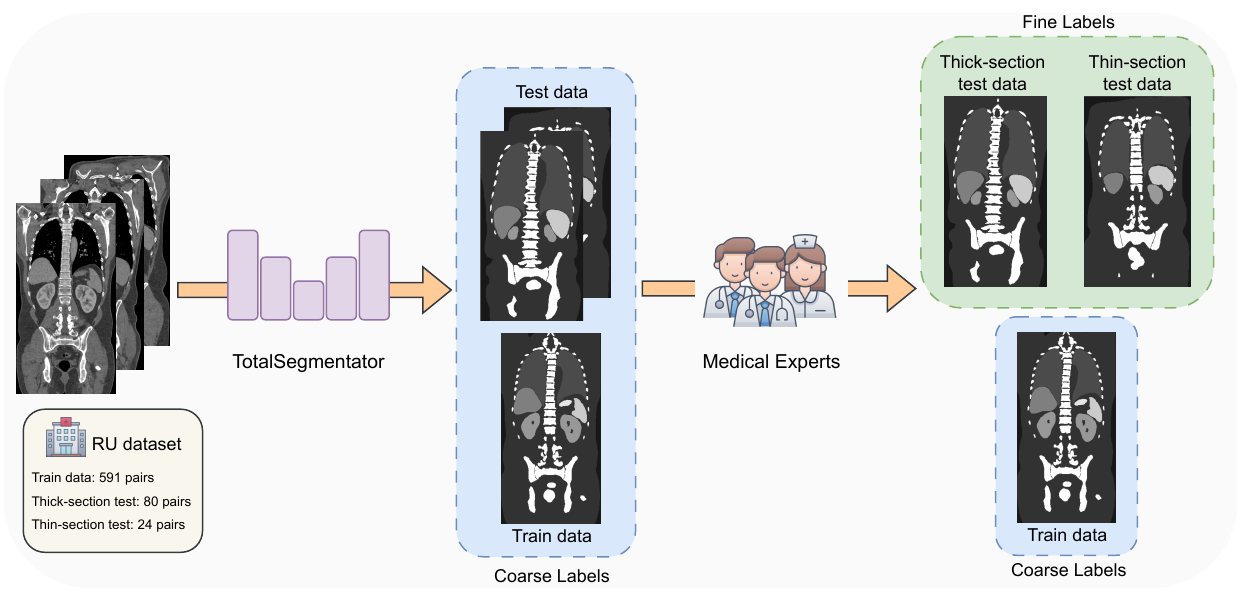}
\end{center}
\caption[readerpipeline] 
{ \label{fig:reader_pipeline} Illustration of the annotation pipeline. We first used TotalSegmentator to obtain the multi-organ coarse labels for all scans in the dataset. Subsequently, medical experts were invited to manually refine the coarse labels of the test set.
}
\end{figure} 

In the initial phase of preprocessing, we examined all scans to verify that they were standardized to the left-posterior-superior (LPS) orientation. Subsequently, we followed a structured pipeline as illustrated in Figure \ref{fig:reader_pipeline} to obtain the annotations for 13 anatomical structures for each CT image. Specifically, the thoracic region includes segmentation masks for the lungs and heart, while the abdominal region encompasses annotations for seven organs: the liver, kidneys, pancreas, spleen, stomach, gallbladder, and prostate. Additionally, the bone region comprises segmentation masks for the pelvis bone, vertebrae, and ribs. A whole-body mask was also delineated for subsequent use. To generate these annotations, we first utilized TotalSegmentator~\citep{Wasserthal23} with its default weights to automatically generate coarse segmentation masks for all 13 target structures across the CT scans in our dataset. While we primarily followed the organ label definitions provided by TotalSegmentator, several modifications were made. Specifically, we merged the vertebrae and sacrum into a single “vertebrae” label; combined the ribs, sternum, and costal cartilages into a unified “ribs” label; grouped the femur and hip under a single “pelvis bone” label; and consolidated the kidney cyst and kidney into one “kidney” label. Additionally, we excluded liver vessels from the “liver” label and retained only the heart chambers excluding the pulmonary artery as the “heart” label. 

To ensure the accuracy of the test set, we conducted a reader study on the Grand Challenge platform\footnote{\url{https://grand-challenge.org/}}, where we uploaded scans from the evaluation dataset and invited eight medical experts, including radiology residents and radiologists from the Netherlands and Vietnam, to manually refine the coarse annotations. The test set was evenly divided into eight parts, with each subset randomly assigned to a reader. Additionally, participants were also required to provide a brief medical report for each case, highlighting any detected abnormalities. Based on these reports, we excluded image pairs in which anatomical structures were inconsistently present across timepoints. Finally, where necessary, two additional radiologists reviewed and verified the refined annotations to ensure optimal quality. 

After obtaining sufficient annotations, we utilized the body segmentation mask to crop the original images to remove any irrelevant information. Depending on slice thickness, the dataset was divided into two subsets: a thin-section subset with a slice spacing of 0.8 mm and a thick-section subset with a slice spacing of either 3 mm or 4 mm. For both subsets, the cropped images were resampled to a uniform resolution of 128x96 pixels with 160 slices. Lastly, to mitigate potential intensity biases caused by various factors such as metal artifacts and inconsistencies in background intensity values, we clipped the intensity range of the dataset to \([-1000, 1000]\) Hounsfield units and normalized to a range of \([0, 1]\).

\subsubsection{External datasets}
External datasets were also resampled to the dimension of 128x96x160. In addition, we followed the final preprocessing step of the in-house dataset to clip the intensity range to \([-1000, 1000]\) and normalize it to the range of \([0, 1]\). 

\subsection{Experiment Setup}
In this study, we solely passed the in-house dataset with the comprehensive body view for training. Specifically, 591 thick-section pairs (1182 for both directions) with a 3-4mm reconstruction interval were employed to train the model. We purposely left out the prostate from the training organ list to become an out-of-distribution organ in the evaluation phase. To enhance the model's performance in cases with large deformations, we further refined the intra-patient models using inter-patient data. This dataset comprises 1,395,942 thick-section pairs, generated by mapping all scans from the original intra-patient dataset. 

Due to a limitation in the computing power and to avoid overfitting, we only fed 400 pairs to the model in each training epoch. Each registration block was trained for 400 epochs using the Adam optimizer with a learning rate of $1 \times 10^{-4}$ and a batch size of 1. On average, training each block took approximately 72 hours on an NVIDIA RTX 3080 Ti GPU with 12GB of VRAM. Since the standalone blocks could be trained in parallel, the total training time for the entire pipeline was reduced to roughly one week. The peak VRAM consumption during training reaches 9.3 GB, while the average VRAM usage remains at 6.5 GB throughout most of each epoch.  %% Add endline

To thoroughly evaluate the proposed method, we used the in-house test set of 104 pairs, including 80 thick-section and 24 thin-section pairs. In this primary experiment, we evaluated the accuracy of TotalRegistrator across 12 anatomical structures, in both intra- and inter-patient scenarios. The inter-patient test set was created by randomly mixing patients from the intra-patient test set, resulting in 200 and 100 pairs for thick-section and thin-section cases, respectively. The results were then compared against two well-known baseline methods: Elastix~\citep{Klein10}, a traditional registration algorithm, and uniGradICON~\citep{Lin24}, a foundation model for image registration. We also conducted several ablation studies, including an investigation of individual component blocks in the TotalRegistrator pipeline to assess the effectiveness of the multi-organ segmentation mask guidance and the "divide-to-conquer" strategy, as well as experiments on the out-of-distribution organ. Additionally, three external datasets, AbdomenCTCT, NLST and H108M, were employed to verify the performance of TotalRegistrator on out-of-distribution data.

To measure the image registration performance of the models in this study, we used the Dice Similarity Coefficient (DSC) metric and the percentage of folding, which are defined as follows.
The DSC is a metric used to measure the spatial overlap between two binary segmentations to evaluate how similar a predicted segmentation is to the ground truth. In image registration specifically, the metric measures how close the segmentation mask of the warped image is compared to the fixed segmentation mask. A DSC of 1 indicates perfect overlap, while a DSC of 0 indicates no overlap. Equation \eqref{DSC} shows the mathematical definition of the DSC metric, where $S_F$ and $S_W$ represent the segmentation masks for the fixed and warped images, respectively. 
\begin{equation}
    \text{DSC}\left(S_F,S_W\right)=\frac{2\left\vert S_{F}\cap (S_{W}) \right\vert}{\left\vert S_F\right\vert+\left\vert S_W\right\vert} \quad,
\label{DSC}
\end{equation}
In image registration, folding occurs when the local deformation inverts the orientation of the space surrounding a voxel, which creates a "fold" and thereby violates the diffeomorphism property. The Jacobian matrix of a 3D deformation field $\Phi(x, y, z)$ is a $3\times3$ matrix of partial derivatives describing local geometric changes. Analyzing its determinant helps assess whether the transformation maintains local invertibility and orientation, which could be defined as:
\begin{equation}
\det(\nabla \Phi(x, y, z)) =
\det \begin{bmatrix}
\frac{\partial \Phi_x}{\partial x} & \frac{\partial \Phi_x}{\partial y} & \frac{\partial \Phi_x}{\partial z} \\
\frac{\partial \Phi_y}{\partial x} & \frac{\partial \Phi_y}{\partial y} & \frac{\partial \Phi_y}{\partial z} \\
\frac{\partial \Phi_z}{\partial x} & \frac{\partial \Phi_z}{\partial y} & \frac{\partial \Phi_z}{\partial z}
\end{bmatrix}
\end{equation}
This determinant provides insight into whether there is a local expansion ($\det{(\nabla{\Phi})} > 1$), a local compression ($\det{(\nabla{\Phi})} < 1$), or a volume preservation ($\det{(\nabla{\Phi})} = 1$). A local inversion typically occurs in regions with negative $\det{(\nabla{\Phi})}$, while $\det{(\nabla{\Phi})} = 0$ indicates the collapse of the local space. The percentage of folding is calculated by the total number of spatially inverted voxels divided by the total number of voxels in the image domain.

\section{Experimental Results} \label{Results}
This section presents the results of a comprehensive evaluation of TotalRegistrator's performance across a wide range of registration scenarios, including longitudinal multi-timepoint, multiphase, intra-patient, inter-patient, in-distribution, and out-of-distribution image registration. Moreover, the multi-center evaluation datasets include substantial variability in patient age, gender, race and CT scanner manufacturers. 

\subsection{In-house dataset}
\subsubsection{Main experiments}

\begin{table*}[!t]
\caption{\\\hspace{\textwidth}Intra-patient registration results on the thick-section test set. This table compares TotalRegistrator with two baseline methods, Elastix and uniGradICON, across 11 anatomical structures. It also evaluates the individual components of the proposed pipeline, including four standalone registration blocks trained independently. For Elastix, we mainly used the default parameter files for both rigid and deformable registration. For uniGradICON, the default model was used without the optional instance optimization. Results are reported as mean~$\pm$~standard deviation. Bold values indicate the highest statistically significant performance for each anatomical structure, determined by a p-value~$<$~0.05.}
\centering
\resizebox{\textwidth}{!}{%
\begin{tabular}{l| c c c| c c c c c}
     \textbf{Organs} & \textbf{Unregistered}& \textbf{Elastix} & \textbf{uniGradICON} & 
     \makecell{\textbf{TotalReg} \\ \textbf{(Thorax)}} & 
     \makecell{\textbf{TotalReg} \\ \textbf{(Abdomen)}} &
     \makecell{\textbf{TotalReg} \\ \textbf{(Bones)}} &
     \makecell{\textbf{TotalReg} \\ \textbf{(Wholebody)}} &
     \textbf{TotalReg} \\
        
\hline \hline &&&&&&&&\\[-0.8em]
Lung & $77.5 \pm 8.3$ & $97.4\pm2.0$ & $\mathbf{97.7 \pm 1.0}$ & 96.4 ± 1.8 & 90.2 ± 5.3 & 92.4 ± 5.1 & 95.2 ± 2.9 & $96.9 \pm 1.5$ \\

Heart & $68.1 \pm 14.7$ & $91.5 \pm 3.1$ & $93.3 \pm 2.1$ & 92.7 ± 5.5 & 84.4 ± 8.3 & 82.7 ± 10.9 & 91.1 ± 6.9 & $93.8 \pm 5.5$ \\

Liver & $72.9 \pm 11.9$ & $92.5 \pm 5.1$ & $94.9 \pm 3.8$ & 87.1 ± 6.8 & 92.8 ± 5.3 & 85.1 ± 8.9 & 93.4 ± 5.0 & $94.9 \pm 4.4$ \\

Kidney & $56.4 \pm 15.0$ & $86.1 \pm 9.2$ & $92.0 \pm 5.7$ & 71.2 ± 10.7 & 88.3 ± 8.0 & 76.5 ± 10.7 & 88.4 ± 7.8 & $92.0 \pm 7.1$ \\

Pancreas & $38.7 \pm 20.5$ & $66.5 \pm 19.1$ & $\mathbf{76.1 \pm 15.6}$ & 48.2 ± 16.9 & 70.3 ± 17.9 & 52.3 ± 17.8 & 71.5 ± 16.8 & $74.7 \pm 17.0$ \\

Spleen & $55.0 \pm 20.8$ & $84.0 \pm 15.5$ & $89.0 \pm 12.7$ & 76.1 ± 15.1 & 88.2 ± 14.9 & 71.3 ± 17.9 & 87.6 ± 15.7 & $\mathbf{90.7 \pm 13.4}$ \\

Stomach & $49.0 \pm 19.4$ & $70.1 \pm 16.3$ & $76.3 \pm 16.0$ & 61.3 ± 16.1 & 76.8 ± 14.8 & 60.9 ± 16.8 & 76.4 ± 15.6 & $\mathbf{79.4 \pm 14.8}$ \\ 

Gallbladder & $20.3 \pm 26.2$ & $50.6 \pm 26.5$ & $63.3 \pm 24.5$ & 33.6 ± 26.2 & 60.8 ± 27.6 & 34.2 ± 27.2 & 63.1 ± 25.4 & $\mathbf{65.5 \pm 25.5}$ \\

Pelvis & $50.4 \pm 15.9$ & $92.6 \pm 4.2$ & $\mathbf{94.2 \pm 3.4}$ & 68.3 ± 13.5 & 70.4 ± 13.2 & 92.3 ± 4.7 & 89.0 ± 6.7 & $93.8 \pm 3.9$ \\

Vertebrae & $50.6 \pm 8.9$ & $\mathbf{88.9 \pm 3.3}$ & $87.7 \pm 2.7$ & 67.2 ± 7.3 & 65.2 ± 8.2 & 84.6 ± 3.3 & 81.7 ± 4.9 & $87.4 \pm 2.5$ \\

Ribs & $16.9 \pm 8.7$ & $75.4 \pm 8.3$ & $73.1 \pm 7.6$ & 47.7 ± 8.9 & 37.5 ± 9.5 & 73.3 ± 5.9 & 67.6 ± 7.7 & $\mathbf{77.3 \pm 5.9}$ \\
        
\hline &&&&&&&&\\[-0.8em]
Avg. DSC & $50.5 \pm 15.5$ & $81.4 \pm 10.2$ & $85.2 \pm 8.6$ & 67.3 ± 11.3 & 72.6 ± 12.2 & 77.6 ± 13.1 & 81.7 ± 11.3 & $\mathbf{86.7 \pm 8.9}$ \\

&&&&&&&&\\[-0.8em]
Avg. $\%$Folding & 0.0 ± 0.0 & 0.2 ± 0.3 & 0.2 ± 0.3 & 0.1 ± 0.1 & 0.1 ± 0.1 & 0.1 ± 0.1 & 0.1 ± 0.1 & 0.2 ± 0.2\\
\hline
\end{tabular}%
}
\label{tab:intra4}
\end{table*}

\begin{figure} [!t]
\begin{center}
\includegraphics[width=1\textwidth]{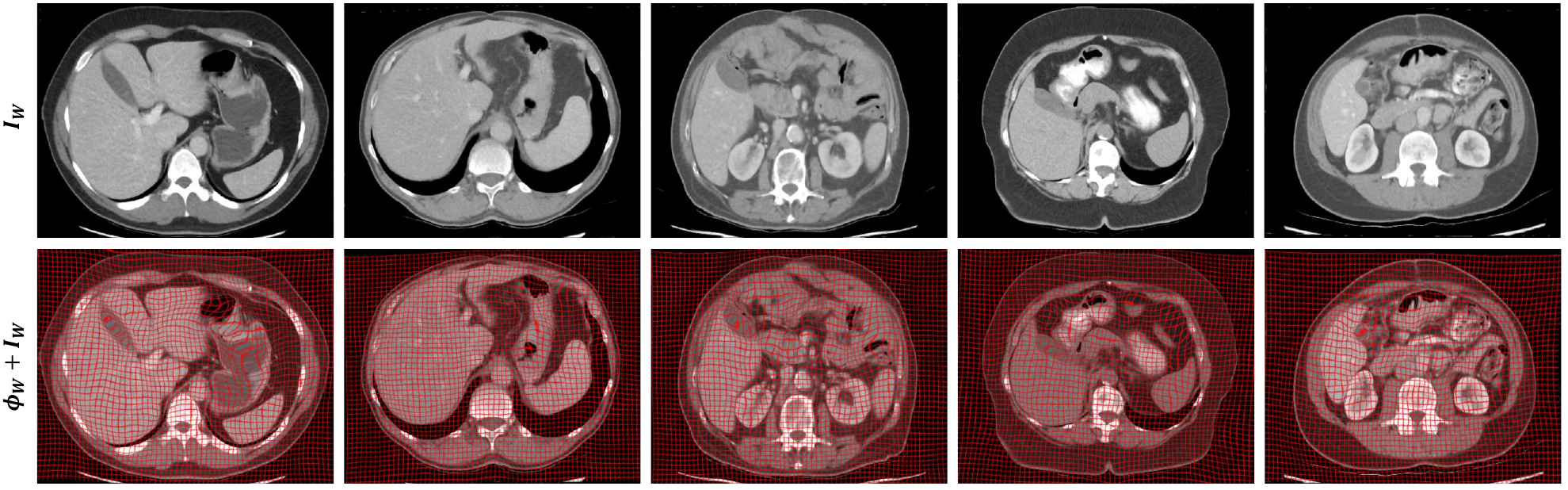}
\end{center}
\caption[folding] 
{ \label{fig:folding} 
Overlay of warped images and their mesh deformation fields. While the overall folding percentage remains low, typical regions such as the gallbladder and stomach often contain large folding errors, either due to their small size or complex deformations. 
}
\end{figure}

\begin{figure} [!t]
\begin{center}
\includegraphics[width=1\textwidth]{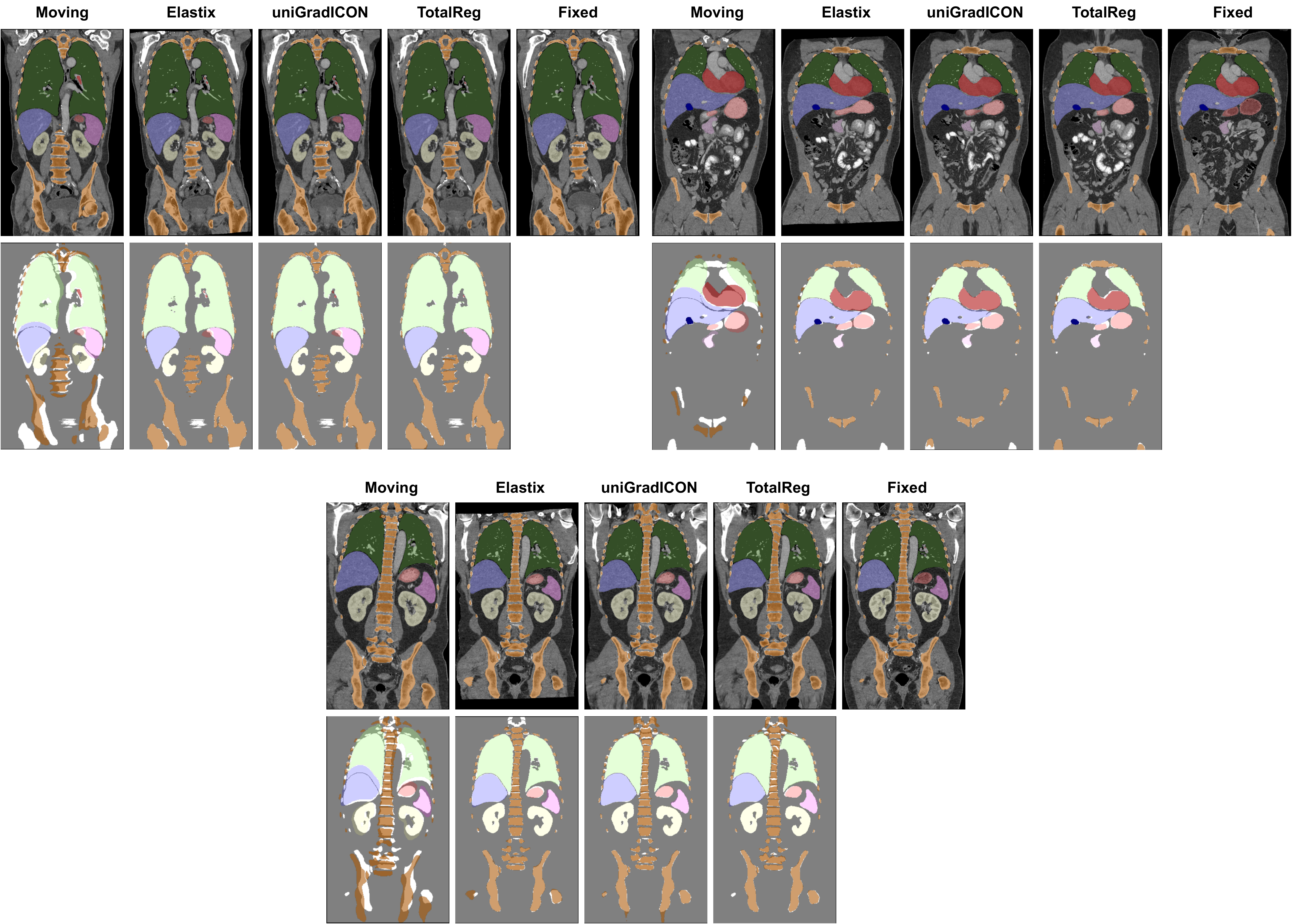}
\end{center}
\caption[summary] 
{ \label{fig:summary} 
Visual examples of the multi-organ, multi-region registration by TotalRegistrator and two baseline methods (Elastix, uniGradICON). The second row in each case shows the multi-organ segmentation masks overlayed between the corresponding warped and fixed images.  
}
\end{figure}

The primary experiments in this study were conducted on the in-house Radboudumc dataset. Results from Table \ref{tab:intra4} show that TotalRegistrator achieves the highest average performance on 11 structure-of-interest, with a notable DSC improvement of 36.7$\%$ compared to unaligned images and 1.5$\%$ compared to the baseline foundation model (uniGradICON). A closer analysis reveals that TotalRegistrator and uniGradICON performed comparably on structures such as the liver, kidneys, and vertebrae. However, uniGradICON outperformed our model on lung and pancreas registration, with leading DSC values of 97.7$\%$ and 76.1$\%$, respectively. In contrast, TotalRegistrator outperformed all baselines on several abdominal organs, including the spleen (90.7$\%$), stomach (79.4$\%$), and gallbladder (65.5$\%$). In this standard intra-patient in-distribution scenario, both TotalRegistrator and uniGradICON maintained low deformation folding rates, with average folding percentages of 0.2$\%$, indicating comparable regularity of the predicted deformation fields (Fig. \ref{fig:folding}). Some visual example cases of this experiment could be seen in Figure \ref{fig:summary}.

\begin{table*}[!t]
\caption{\\\hspace{\textwidth}Inter-patient registration results on the thick-section test set. This table compares TotalRegistrator with two baseline methods, Elastix and uniGradICON, across 11 anatomical structures. It also evaluates the individual components of the proposed pipeline, including four standalone registration blocks trained independently. Results are reported as mean~$\pm$~standard deviation. Bold values indicate the highest statistically significant performance for each anatomical structure, determined by a p-value~$<$~0.05.} 
\centering
\resizebox{\textwidth}{!}{%
\begin{tabular}{l| c c c| c c c c c}
     \textbf{Organs} & \textbf{Unregistered}& \textbf{Elastix} & \textbf{uniGradICON} & 
     \makecell{\textbf{TotalReg} \\ \textbf{(Thorax)}} & 
     \makecell{\textbf{TotalReg} \\ \textbf{(Abdomen)}} &
     \makecell{\textbf{TotalReg} \\ \textbf{(Bones)}} &
     \makecell{\textbf{TotalReg} \\ \textbf{(Wholebody)}} &
     \textbf{TotalReg} \\
        
         \hline \hline &&&&&&&&\\[-0.8em]
         Lung & $65.5\pm10.2$  & $88.8\pm14.1$ & ${\mathbf{96.5\pm1.8}}$ &  $92.8\pm7.1$ &  $75.0\pm8.9$ &  $79.5\pm6.6$ &  $88.9\pm5.7$ &  $91.3\pm4.3$ \\
         
         Heart & $48.3\pm18.8$ & $74.0\pm18.6$ & $84.5\pm7.2$ &  $88.2\pm6.2$ &  $63.5\pm10.2$ &  $48.5\pm8.3$ &  $81.5\pm7.9$ &  $85.0\pm5.7$ \\
         
         Liver & $49.9\pm17.1$ & $68.4\pm18.5$ & $81.4\pm10.1$ &  $66.9\pm8.6$ &  $85.9\pm7.3$ &  $52.2\pm9.4$ &  $83.5\pm6.9$ &  ${\mathbf{84.8\pm7.0}}$ \\
        
        Kidney & $28.4\pm13.8$ & $38.7\pm18.9$ & $62.3\pm19.9$ &  $31.7\pm9.3$ &  $69.1\pm12.1$ &  $32.6\pm7.4$ &  $65.8\pm7.2$ &  ${\mathbf{66.2\pm6.3}}$ \\
        
        Pancreas & $13.5\pm12.1$ & $21.1\pm16.4$ & $34.7\pm18.4$ &  $17.2\pm9.2$ &  $43.1\pm14.4$ &  $14.0\pm8.3$ &  $44.5\pm7.3$ &  ${\mathbf{45.7\pm9.5}}$ \\
        
        Spleen & $26.0\pm19.3$ & $47.2\pm23.3$ & $62.8\pm20.9$ &  $47.1\pm10.8$ &  $78.6\pm11.8$ &  $31.9\pm11.5$ &  $75.6\pm8.0$ &  ${\mathbf{74.5\pm9.0}}$ \\
        
        Stomach & $20.7\pm14.9$ & $31.5\pm18.5$ & $40.8\pm19.7$ &  $27.8\pm10.3$ &  $57.9\pm12.1$ &  $20.5\pm9.7$ &  $54.7\pm8.3$ &  ${\mathbf{54.4\pm7.6}}$ \\ 
        
        Gallbladder & $2.0\pm5.9$ & $7.6\pm13.4$ & $13.5\pm17.8$ &   $4.3\pm4.5$ &  $25.9\pm9.7$ &   $3.4\pm3.7$ &  $27.0\pm9.0$ &  ${\mathbf{28.8\pm7.7}}$ \\

        Pelvis & $36.0\pm11.8$ & $65.9\pm20.1$ & ${\mathbf{89.8\pm8.4}}$ &  $33.0\pm8.7$ &  $37.2\pm8.1$ &  $85.2\pm10.4$ &  $74.0\pm9.3$ &  $81.3\pm8.1$ \\

        Vertebrae & $38.6\pm8.1$ & $63.0\pm15.0$ & ${\mathbf{76.2\pm4.9}}$ &  $45.8\pm10.6$ &  $43.8\pm7.8$ &  $74.7\pm8.1$ &  $65.9\pm8.6$ &  $70.5\pm7.2$ \\
        
        Ribs & $9.0\pm4.8$ & $36.7\pm9.5$ & ${\mathbf{50.2\pm5.3}}$ &  $28.2\pm6.7$ &  $13.1\pm4.5$ &  $56.7\pm9.8$ &  $38.4\pm7.1$ &  $44.7\pm8.6$ \\
        \hline &&&&&&&&\\[-0.8em]
        Avg. DSC& $30.7\pm12.4$ & $49.3\pm16.9$ & $63.0\pm12.2$ &  $43.1\pm13.0$ &  $53.9\pm15.5$ &  $44.0\pm14.1$ &  $62.2\pm9.4$ &  $\mathbf{66.1\pm8.3}$ \\
        
         &&&&&&&&\\[-0.8em]
        Avg. $\%$Folding & 0.0 ± 0.0 & 1.9 ± 1.8 & $0.3 \pm 0.3$ & 0.1 ± 0.4 & 0.5 ± 0.9 & 0.2 ± 0.5 & 0.2 ± 0.2 & 0.2 ± 0.4 \\
        \hline
    \end{tabular}%
    }
\label{tab:inter4}
\end{table*}

Table \ref{tab:inter4} presents results on the more challenging inter-patient scenario, which involves substantial anatomical variability and complex deformations across different individuals. TotalRegistrator achieved the highest average DSC among all methods, but performance differences across specific anatomical structures were more pronounced than in the intra-patient setting. Specifically, the proposed method outperforms other baseline methods in all abdominal organs with a large margin. Notably, certain organs such as the pancreas, spleen, stomach and gallbladder exhibit an improvement of more than 10$\%$ DSC compared to the baseline foundation model. However, uniGradICON obtains a superior performance in lung and other bones registration with a significant difference of more than 5$\%$ DSC compared to the proposed method. Despite the increased complexity, TotalRegistrator maintained low deformation irregularity, with an average folding rate of 0.2$\%$, while uniGradICON exhibited a slightly higher folding rate of 0.3$\%$. 

\begin{figure} [!t]
\begin{center}
\includegraphics[width=1\textwidth]{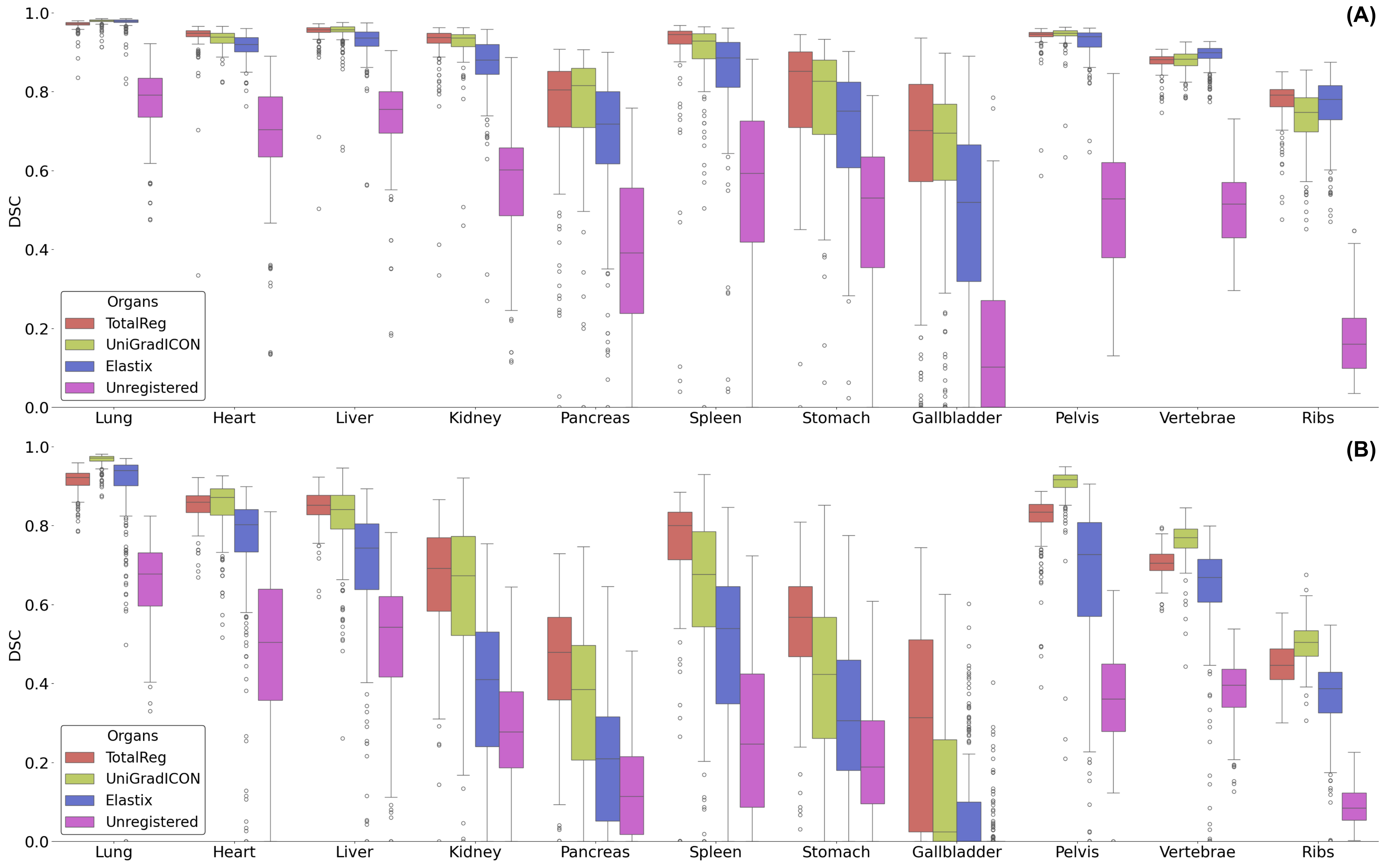}
\end{center}
\caption[boxplotDSC] 
{ \label{fig:plotdsc} 
Boxplot representation of the intra (A) and inter-patient (B) registration results on the thick-section test set. In general, Totalreg shows better stability with more centralized boxes. In addition, all methods have the most performance variability in small or largely deformed abdominal organs such as the gallbladder, pancreas or stomach.    
}
\end{figure} 

To assess the impact of slice thickness and interpolation on the image registration performance, we conducted additional experiments on the thin-section test set with a 0.8mm slice reconstruction interval. Tables \ref{tab:intra08} and \ref{tab:inter08} show that the overall trends are consistent with those observed on the thick-section test set. In both intra and inter patient scenarios, TotalRegistrator achieves the highest average DSC values of 87.9$\%$ and 67.3$\%$ respectively. At the per-structure level, the proposed method consistently outperforms other baselines on abdominal organs with a clearer difference compared to previous results on the thick-section test set. Meanwhile, uniGradICON achieved the highest DSC values for lung registration across all scenarios, and higher scores for bone structures in inter-patient experiments compared to TotalRegistrator.       

\begin{table*}[!t]
\caption{\\\hspace{\textwidth}Intra-patient registration results on the thin-section test set. This table compares TotalRegistrator with two baseline methods, Elastix and uniGradICON, across 11 anatomical structures. Results are reported as mean~$\pm$~standard deviation. Bold values indicate the highest statistically significant performance for each anatomical structure, determined by a p-value~$<$~0.05.} 
\centering
\resizebox{0.55\textwidth}{!}{%
\begin{tabular}{l| c c c| c}
     % & 
    % \multicolumn{4}{c|}{\textbf{SLIVER}} &  
    % \multicolumn{3}{c|}{\textbf{H108CE}} \\ &&&&&&&&\\[-0.8em]
     \textbf{Organs} & \textbf{Unregistered} & \textbf{Elastix} & \textbf{uniGradICON} & 
     \textbf{TotalReg} \\
        
         \hline \hline &&&&\\[-0.8em]
         Lung & $76.0\pm10.2$ & $97.9\pm0.8$ & $\mathbf{98.1\pm0.4}$ & $97.3\pm0.6$ \\
         
         Heart & $66.7\pm14.9$ & $93.0\pm2.0$ & $94.6\pm1.4$ & $\mathbf{95.7\pm1.0}$ \\
         
         Liver & $71.5\pm11.2$ & $93.8\pm3.1$& $95.8\pm2.2$ & $\mathbf{96.0\pm1.9}$ \\
        
        Kidney & $58.8\pm15.5$ & $88.8\pm5.4$ & $93.8\pm2.1$ & $\mathbf{94.3\pm1.8}$ \\
        
        Pancreas & $35.4\pm20.8$ & $67.8\pm17.1$ & $\mathbf{79.1\pm10.3}$ & $76.7\pm13.0$ \\
        
        Spleen & $58.8\pm18.4$ & $84.4\pm13.6$ & $90.4\pm8.6$ & $\mathbf{92.5\pm9.4}$ \\
        
        Stomach & $47.3\pm21.4$ & $71.1\pm15.6$ & $78.2\pm14.5$ & $\mathbf{81.7\pm11.0}$ \\ 
        
        Gallbladder & $25.5\pm32.0$ & $52.7\pm28.0$ & $66.2\pm23.4$ & $\mathbf{69.2\pm24.9}$ \\

        Pelvis & $54.3\pm15.8$ & $93.7\pm3.0$ & $95.2\pm1.1$ & $95.3\pm0.9$ \\

        Vertebrae & $49.6\pm9.0$ & $\mathbf{91.0\pm2.1}$ & $87.9\pm 2.4$ & $88.6\pm1.8$ \\
        
        Ribs & $17.3\pm10.7$ & $80.0\pm7.4$ & $74.3\pm7.6$ & $80.0\pm5.0$ \\
        \hline &&&&\\[-0.8em]
        Avg DSC & $51.0 \pm 16.3$ & $83.1 \pm 8.9$ & $86.7\pm6.7$ & $\mathbf{87.9\pm6.5}$\\
        &&&&\\[-0.8em]
        Avg $\%$Folding & 0.0 ± 0.0 & 0.0 ± 0.0 & 0.1 ± 0.2 & 0.5 ± 0.4\\
        \hline
        
    \end{tabular}%
    }
\label{tab:intra08}
\end{table*}

\subsubsection{Ablation studies}
In addition to the primary experiments, we conducted a series of ablation studies to assess the individual contributions of different components in the TotalRegistrator pipeline. First, we investigated the performance of each standalone registration block and quantified its contribution and influence in the pipeline. As shown in Tables \ref{tab:intra4} and \ref{tab:inter4}, each block performed best on its corresponding anatomical region. Specifically, out of the three components, the thorax block achieved the highest DSC values of 96.4$\%$ (intra-patient) and 92.8$\%$ (inter-patient) for the lungs, and 92.7$\%$ and 88.2$\%$ for the heart. Similarly, the abdomen and bone blocks achieved the highest DSCs for abdominal organs and skeletal structures, respectively. 

Second, we evaluated the standalone performance of the whole-body registration block. This block achieved average DSC values of 81.7$\%$ and 62.2$\%$ in the intra- and inter-patient settings, respectively, and produced consistent results across all anatomical regions. However, in region-specific tasks, it was outperformed by the corresponding standalone blocks. For example, in the inter-patient setting, the ribs achieved a DSC of 56.7$\%$ in the bone-specific block compared to 38.4$\%$ in the whole-body block, and heart registration showed a DSC of 88.2$\%$ in the thorax block versus 81.5$\%$ in the whole-body block.

% Second, the standalone Wholebody block demonstrates the effectiveness of incorporating segmentation masks in training multi-organ registration models. However, despite having promising results across all anatomical structures and a superior average DSC of 81.7$\%$ and 62.2$\%$, it tends to be surpassed in the region-of-focus of each corresponding standalone block. For instance, there is a large difference of 18.3$\%$ DSC in inter-patient ribs registration between the standalone Wholebody block and the standalone Bones block, or 6.7$\%$ DSC difference in inter-patient heart registration between the Wholebody and Thorax block. This limitation demonstrates the importance of the "divide-and-conquer" idea in this study. %% Remember to move the interpretation here to either Experiment Setup or Discussion + Revise tense

Lastly, we evaluated the generalizability of TotalRegistrator on an out-of-distribution organ. For a fair comparison, we selected the prostate because it was also excluded from the training data of uniGradICON. According to table \ref{tab:OOD}, all models exhibit significant improvements from no-alignment results across both experiment scenarios. The baseline foundation model achieves DSCs of $61.5\%$ and $61.3\%$ for this organ, while TotalRegistrator follows with DSCs of $58.0\%$ and $57.0\%$, respectively.  

\begin{table*}[!t]
\caption{\\\hspace{\textwidth}Inter-patient registration results on the thin-section test set. This table compares TotalRegistrator with two baseline methods, Elastix and uniGradICON, across 11 anatomical structures. Results are reported as mean~$\pm$~standard deviation. Bold values indicate the highest statistically significant performance for each anatomical structure, determined by a p-value~$<$~0.05.} 
\centering
\resizebox{0.55\textwidth}{!}{%
\begin{tabular}{l| c c c| c}
     % & 
    % \multicolumn{4}{c|}{\textbf{SLIVER}} &  
    % \multicolumn{3}{c|}{\textbf{H108CE}} \\ &&&&&&&&\\[-0.8em]
     \textbf{Organs} & \textbf{Unregistered} & \textbf{Elastix} & \textbf{uniGradICON} & 
     \textbf{TotalReg} \\
        
         \hline \hline &&&&\\[-0.8em]
         Lung & $65.5\pm9.8$ & $89.5\pm12.1$ & $\mathbf{97.0\pm1.1}$ & $91.3\pm2.9$ \\
         
         Heart & $44.3\pm19.9$ & $73.8\pm18.1$ & $84.8\pm7.9$ & $\mathbf{86.1\pm3.7}$ \\
         
         Liver & $49.6\pm15.5$ & $69.1\pm15.9$& $82.3\pm8.3$ & $\mathbf{85.2\pm3.7}$ \\
        
        Kidney & $29.5\pm13.0$ & $42.9\pm17.0$ & $67.8\pm16.7$ & $\mathbf{70.7\pm11.1}$ \\
        
        Pancreas & $13.2\pm13.3$ & $20.4\pm14.1$ & $33.5\pm17.9$ & $\mathbf{46.0\pm13.4}$ \\
        
        Spleen & $28.4\pm18.4$ & $51.1\pm20.6$ & $67.6\pm16.7$ & $\mathbf{79.0\pm9.8}$ \\
        
        Stomach & $20.3\pm15.6$ & $33.8\pm19.0$ & $42.9\pm20.1$ & $\mathbf{59.2\pm14.9}$ \\ 
        
        Gallbladder & $2.6\pm8.3$ & $7.3\pm14.5$ & $12.3\pm20.0$ & $\mathbf{26.7\pm26.1}$ \\

        Pelvis & $37.6\pm10.9$ & $64.4\pm18.1$ & $\mathbf{91.9\pm2.2}$ & $83.3\pm3.8$ \\

        Vertebrae & $37.7\pm7.6$ & $59.7\pm12.7$ & $\mathbf{74.8\pm3.6}$ & $68.7\pm3.3$ \\
        
        Ribs & $8.7\pm4.7$ & $36.9\pm10.2$ & $\mathbf{50.8\pm5.4}$ & $44.0\pm5.9$ \\
        \hline &&&&\\[-0.8em]
        Average & $30.7 \pm 12.5$ & $50.0 \pm 15.7$ & $64.2\pm10.9$ & $\mathbf{67.3\pm9.0}$\\
        &&&&\\[-0.8em]
        Avg $\%$Folding & 0.0 ± 0.0 & 0.0 ± 0.0 & 0.2 ± 0.3 & 0.7 ± 0.7\\
        \hline
    \end{tabular}%
    }
\label{tab:inter08}
\end{table*}

\subsection{External datasets}

\begin{table*}[!t]
\caption{\\\hspace{\textwidth}Out-of-distribution organ experiment. This table compares TotalRegistrator with two baseline methods, Elastix and uniGradICON, on the prostate. Results are reported as mean ± standard deviation.}
\centering
\resizebox{0.72\textwidth}{!}{%
\begin{tabular}{l| c c c c}
     \textbf{Data} & \textbf{Unregistered} & \textbf{Elastix} & \textbf{uniGradICON} & \makecell{\textbf{TotalReg}} \\
     \hline \hline &&&&\\[-0.8em]
     Intra-patient registration - 3/4mm test set  & $34.4\pm26.1$ & 61.5 ± 11.3 & $61.5\pm 30.4$ & $58.0\pm27.1$ \\
     % Inter-patient registration - 3/4mm test set  & $25.9\pm40.6$ & 28.8 ± 11.3 & $\mathbf{34.0\pm41.9}$ & $31.9\pm40.9$ \\
     Intra-patient registration - 0.8mm test set& $33.9\pm29.3$ & 60.5 ± 25.8 & $61.3\pm25.4$  & $57.0\pm24.3$ \\
     % Inter-patient registration - 0.8mm test set & $21.2\pm37.2$ & 28.5 ± 9.0 & $\mathbf{30.2\pm38.3}$ & $28.2\pm37.7$ \\
     \hline
\end{tabular}%
}
\label{tab:OOD}
\end{table*}

\begin{table}[!t]
\caption{\\\hspace{\textwidth}Ablation study on the multiphase liver H108M dataset. This table compares TotalRegistrator with six other state-of-the-art baselines. Results of baselines were reported in~\cite{Pham24}.}
    \centering
    \resizebox{0.55\textwidth}{!}{%
    \begin{tabular}{c|ccccccc}
        \toprule
        Metric & \textbf{ANTs} & \textbf{Elastix} & \textbf{VoxelMorph} & \textbf{DPRN} & \textbf{VTN} & \textbf{CMAN} & \textbf{TotalReg} \\
        \midrule
        DSC & 95.0 & 94.2 & 94.0 & 92.1 & 93.6 & 94.0 & 91.9 \\
        \bottomrule
    \end{tabular}%
    }
    \label{tab:h108}
\end{table}

\begin{table*}[!t]
\caption{\\\hspace{\textwidth}Ablation study on the abdominal AbdomenCTCT and lung NLST datasets. This table compares TotalRegistrator with two baseline methods: SyN and uniGradICON. Results of baselines in AbdomenCTCT were reported in~\cite{Lin24}.} 
\centering
\resizebox{0.66\textwidth}{!}{%
\begin{tabular}{c|c c c c c }
     \textbf{Datasets} & \textbf{Unregistered} & \textbf{SyN/Elastix} & 
     \textbf{uniGradICON} & 
     \makecell{\textbf{uniGradICON} \\ \textbf{(excl. Abdomen)}} & 
     % \makecell{\textbf{uniGradICON} \\ \textbf{(wo Abdomen IO)}} &
     \textbf{TotalReg} \\
     \hline \hline &&&&&\\[-0.8em]
     AbdomenCTCT  &  25.90 & 25.20 & 48.30 & 34.10 & 36.20\\
     \hline &&&&&\\[-0.8em]
     NLST       & 90.88 & 97.01 & 98.59 & N/A & 97.40\\
     \hline
\end{tabular}%
}
\label{tab:l2r}
\end{table*}

\begin{figure} [!t]
\begin{center}
\includegraphics[width=0.35\textwidth]{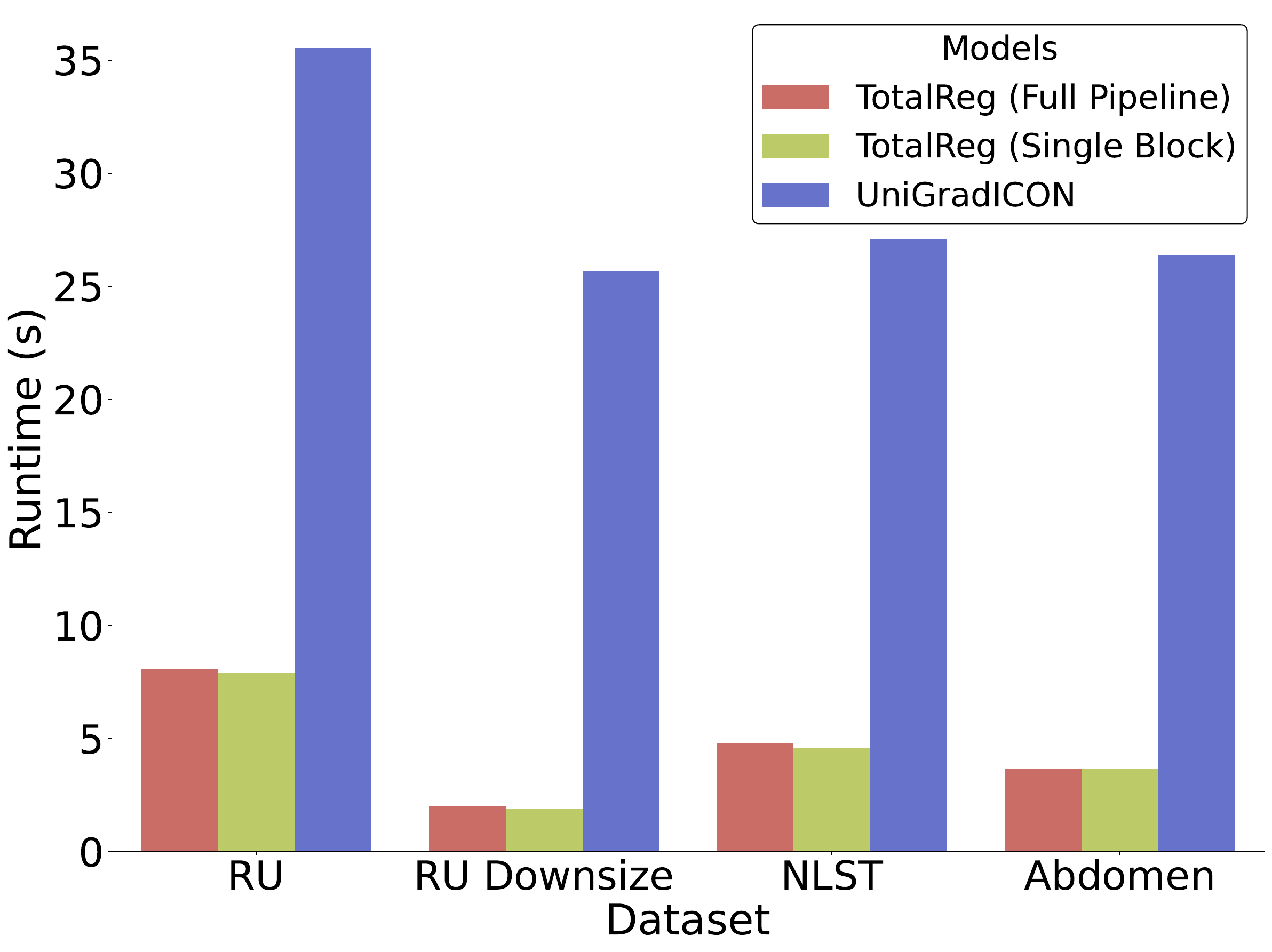}
\end{center}
\caption[runtime] 
{ \label{fig:runtime} 
Runtime comparison between TotalReg and uniGradICON across three datasets. In general, TotalReg has faster funtime than uniGradICON, while there is no distinguishable difference in runtime between single and multiple registration blocks of TotalReg. Notably, in the downsizing of 128x96x160 used in this study, the mean inference time for TotalReg is less than two seconds.  
}
\end{figure}
\begin{figure} [!t]
\begin{center}
\includegraphics[width=1\textwidth]{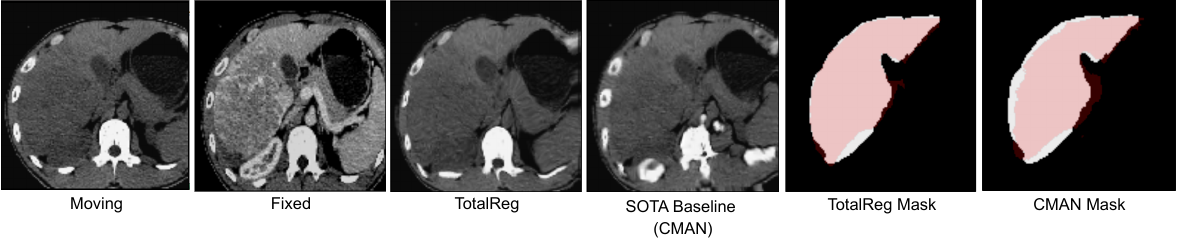}
\end{center}
\caption[H108M] 
{ \label{fig:cman_totalreg} 
Example of a case in the multiphase liver CT dataset H108M. The moving image is a non-contrast scan with indistinguishable boundaries among anatomical structures, while the fixed image is contrast enhanced with clear visualization of anatomies. From the warped images and segmentation masks overlay, the baseline organ-specific model struggles to delineate adjacent anatomies from each other. Meanwhile, Totalreg shows better performance separating the liver from surrounding organs and structures.  
}
\end{figure}

 
To assess the generalizability of TotalRegistrator on unseen data from external institutions, we collected two popular datasets from the Learn2Reg challenge and another dataset from our collaborating hospital in Asia. Table \ref{tab:h108} shows experiment results on the liver multiphase CT dataset. Specifically, TotalRegistrator performs on par with a baseline model (DPRN) but lags slightly behind other methods by roughly 2$\%$ DSC. Visual inspection of the warped images (Fig. \ref{fig:cman_totalreg}) showed that, although TotalRegistrator yielded slightly lower DSC values on the liver dataset, it produced smoother and more spatially coherent deformations around the liver region compared to other methods. Table \ref{tab:l2r} contains experiment results on the AbdomenCTCT and NLST datasets. Regarding the abdominal data, TotalRegistrator outperforms uniGradICON with a DSC value of 36.20$\%$ when AbdomenCTCT remains unseen by both models. However, when uniGradICON was trained on this dataset, its performance significantly increased to 48.3$\%$. On the NLST dataset, uniGradICON obtained a DSC of 98.59$\%$, followed by TotalRegistrator with 97.40$\%$.          

\section{Discussion} \label{Discussion}
This section provides an analysis and interpretation of the experimental results presented in the previous section. These findings reinforce the relevance and significance of the contributions outlined in Section \ref{Intro}, including a lightweight, general-purpose registration model built upon a single well-curated dataset and a field decomposition strategy.

\subsection{In-house dataset}
% \textcolor{red}{Tab:Downscale: Affect of scale on the overall performance}

The experiments on the in-house dataset showed consistent trends across both intra- and inter-patient registration settings (Fig.~\ref{fig:plotdsc}). TotalRegistrator obtains the highest average DSC in both intra- and inter-patient registration scenarios. Furthermore, the proposed method demonstrates better robustness in the abdomen region compared to other baselines, with a more obvious dominance in cases with highly complex deformations. This finding confirms the potential of a single, well-curated dataset in the training of a multi-organ registration model. In contrast, uniGradICON was trained on a mixture of public datasets with an imbalanced representation across anatomical regions. For instance, its lung training set included 899 intra-patient scan pairs, while the abdominal training data comprised only 30 scans~\citep{Lin24}. This imbalance may have contributed to the observed performance gap between the two models in different anatomical regions, with uniGradICON achieving higher scores in lung registration, and TotalRegistrator performing better in the abdomen.

Evaluation on the thin-section test set further confirmed that the model’s performance was not sensitive to reconstruction thickness. Although TotalRegistrator was trained on downsampled volumes with a fixed resolution of 128×96×160, it generalized well to high-resolution test scans containing more than 500 slices. This indicates that the model’s compact input size is sufficient for learning robust deformations, while also supporting efficient training and inference on standard 11GB GPUs. In comparison, uniGradICON operates on 175×175×175 volumes, requiring significantly more memory and runtime (Fig.~\ref{fig:runtime}). 

The ablation studies on the in-house dataset further demonstrate the rationale behind the proposed field decomposition approach in this work. The standalone Wholebody block achieved good average DSC across all organs, but was outperformed by the region-specific blocks in their respective anatomical domains. This finding suggests that a minimum UNet model can already learn to align multiple anatomical structures simultaneously using a single well-curated whole-body CT dataset and multi-organ segmentation masks. However, the model still struggles with the parallel optimization of various deformation types from multiple anatomical structures, even with sufficient support from multi-organ segmentation masks. This limitation, therefore, facilitates the need to decompose one complex deformation field into multiple smaller ones based on both regions and tasks. Additionally, instead of independently training the Wholebody block like other blocks in the pipeline, the strategy of adapting it to the outputs from preceding blocks also indicates its effectiveness in handling the folding exploding issue while still ensuring a smooth combination of other blocks.

\subsection{External datasets}

The experiment on the liver multiphase dataset reveals the potential of TotalRegistrator on unseen tasks and data. Despite not being trained on non-contrast CT scans, the model is still capable of accurately identifying and aligning multiple abdominal organs, particularly the liver in this case. Its robustness to varying contrast levels highlights the effectiveness of the Mutual Information-based similarity loss used during training. Furthermore, TotalRegistrator demonstrates strong generalizability across racial demographics, achieving good performance on an Asian dataset despite being trained exclusively on a European dataset. 

Meanwhile, the other ablation studies on two Learn2Reg datasets further consolidate the key findings obtained from primary experiments. It can be seen that uniGradICON consistently leads in the lung registration task, while TotalRegistrator tends to perform better in the abdominal region. However, for inter-patient test sets with severe initial misalignments, such as those in the AbdomenCTCT dataset, a refinement using the training set can enhance the registration performance.       

\subsection{Limitations}
Despite its potential, this study still has several limitations. First, TotalRegistrator exhibits a promising yet rather unstable performance in bone registration. This is due to the deformable component in the Bones registration block in the pipeline. In future studies, constraining it to only local rigid transformations based on accurate segmentation masks guidance may improve the performance. Second, while the overall folding percentage remains low, typical regions such as the gallbladder and stomach often contain large folding errors, either due to their small size or complex deformations. This issue could be resolved by customizing a specific regularization weight for each organ. Third, the multi-organ registration in this work is mainly guided by segmentation masks via the DSC metric, making it biased towards aligning only the global shapes and unaware of the local details inside the organs. Therefore, incorporating local details-aware metrics such as keypoint distance may improve the results. Fourth, the generalizability on out-of-distribution organs should be extensively evaluated on more organs of different shapes and sizes, instead of only the prostate. Lastly, we only utilized the minimum architecture UNet and losses to highlight the field decomposition approach and its ease of implementation. Thus, more advanced architectures and losses can be deployed to further improve the performance of TotalRegistrator.

\section{Conclusion}    
In this study, we have proposed a novel strategy for the development of a general-purpose whole-body CT image registration model. First, we constructed a large-scale longitudinal intra-patient whole-body CT dataset, ensuring a balanced distribution of anatomical structures across scans. This approach eliminates the need for multiple structure-specific datasets, while retaining sufficient anatomical variability for effective model training. Second, we introduced a multi-cascade region-specific field decomposition approach to address the challenges posed by complex deformations in multi-organ registration. Finally, our model was optimized for reimplementation on standard GPUs with 11GB of VRAM, making it easily accessible to the research community. The proposed method has been rigorously evaluated through multiple scenarios, such as inter/intra-patient registration, longitudinal/multiphase registration, and in-house/out-of-distribution registration. Experimental results demonstrated that TotalRegistrator achieved comparable performance with the state-of-the-art image registration foundation model on whole-body CT image registration while requiring significantly fewer computing resources. In future studies, our method can be fully extended to all anatomical structures and modalities provided by TotalSegmentator to become a robust and lightweight foundation model for image registration. With its versatility and ease of implementation, TotalRegistrator has the potential to serve as an effective baseline method for rapid pilot studies in research and certain clinical settings.   

\section*{Acknowledgments}
This research is funded by the Dutch Research Council (NWO) under grant number Veni-21121 for Applied and Engineering Sciences (Spotting the Differences: AI-based change detection in medical images).

\section*{Declaration of Competing Interest}
The authors have no competing interests to declare.

\section*{Declaration of generative AI and AI-assisted technologies in the writing process}
During the preparation of this work, the authors used ChatGPT in order to check for grammatical errors and improve the clarity of the manuscript. After using this tool/service, the authors reviewed and edited the content as needed and take full responsibility for the content of the publication.

%\section*{\itshape Reference style}

%%Harvard
\bibliographystyle{model2-names.bst}\biboptions{authoryear}
\bibliography{refs}

% \newpage
% \appendix
% \section*{APPENDIX}\label{appendix}

\end{document}